\documentclass[onecolumn,final,traditabstract]{aa}
\pdfoutput=1    %%% force arxiv to use pdflatex
\usepackage{url}
\usepackage{amsmath}
\usepackage{graphicx}
\usepackage{txfonts}
\usepackage[nonamebreak,authoryear]{natbib} 
\bibpunct{(}{)}{;}{a}{}{,} % to follow the A&A style
\usepackage{ifthen}%required by natbib and A&A
\usepackage[breaklinks,colorlinks,citecolor=blue,backref=section]{hyperref}
\usepackage[usenames,dvipsnames]{color}

\makeatletter
\if@onecolumn
        \newcommand{\cols}{1}% onecolumn
\else
        \newcommand{\cols}{2}% twocolumn
\fi
\makeatother
\newcommand{\onetwocols}[3]{%\onetwocols{\cols}{1_col}{2_col}
   \ifthenelse{\equal{#1}{1}}{#2}{#3}
}

\newcommand{\matA}{{\bf A}}
\newcommand{\matB}{{\bf B}}
\newcommand{\matC}{{\bf C}}
\newcommand{\matD}{{\bf D}}
\newcommand{\matH}{{\bf H}}
\newcommand{\matJ}{{\bf J}}

\newcommand{\matN}{{\bf N}}
\newcommand{\matR}{{\bf R}}
\newcommand{\matX}{{\bf X}}
\newcommand{\matY}{{\bf Y}}
\newcommand{\vecd}{{\bf d}}
\newcommand{\vece}{{\bf e}}

\newcommand{\vecn}{{\bf n}}
\newcommand{\vecr}{{\bf r}}

\newcommand{\vectortwo}[2]%
{\begin{pmatrix} #1\\#2\end{pmatrix}}
\newcommand{\vectorthree}[3]%
{\begin{pmatrix} #1\\#2\\#3\end{pmatrix}}
\newcommand{\matrixthree}[9]%
{\begin{pmatrix}#1 & #2 & #3 \\#4 & #5 & #6 \\#7 & #8 & #9\end{pmatrix}}
\newcommand{\vectorfour}[4]%
{\begin{pmatrix} #1\\#2\\#3\\#4\end{pmatrix}}
\newcommand{\vectorsix}[6]%
{\begin{pmatrix} #1\\#2\\#3\\#4\\#5\\#6\end{pmatrix}}
\newcommand{\rowfour}[4]%
{{ #1 & #2 & #3  & #4
}}

\newcommand{\trace}{{\rm Tr}}

\newcommand{\wjjj}[6]%  3j-Wigner Symbol
{\begin{pmatrix} #1 & #2 & #3 \\#4 & #5 & #6 \end{pmatrix}}
\newcommand{\jm}[4]%  2x2 Jones Matrix
{\begin{pmatrix} #1 & #2 \\#3 & #4 \end{pmatrix}}

\newcommand{\VEV}[1]{\ensuremath{\left\langle{#1}\right\rangle}}
\newcommand{\hatb}{\hat{b}}
\newcommand{\pC}{\widetilde{C}}
\newcommand{\hC}{\hat{C}}
\newcommand{\tvm}{\widetilde{{\bf m}}}
\newcommand{\tm}{\widetilde{m}}

\newcommand{\tomega}{\widetilde{\omega}}
\newcommand{\tOmega}{\widetilde{\Omega}}
\newcommand{\pO} {\hat{\Omega}^{s}}
\newcommand{\pOn}{\hat{\Omega}^{-s}}
\newcommand{\pOs}{\hat{\Omega}^{s*}}

\newcommand{\cpe}{\rho}
\newcommand{\cpem}[1]{\cpe_{#1}}  % model cross-polar-efficiency
\newcommand{\cpei}[1]{\cpe_{#1}'} % actual CPE
\newcommand{\myxi}{\xi} % elements of Omega for ideal scanning
\newcommand{\fwhm}{\theta_{\rm FWHM}}
\newcommand{\bI}{\widetilde{I}}
\newcommand{\bQ}{\widetilde{Q}}
\newcommand{\bS}{\widetilde{S}}
\newcommand{\bU}{\widetilde{U}}
\newcommand{\bV}{\widetilde{V}}

\newcommand{\mydot}{\ensuremath{\, . \, }}
\newcommand{\lmax}{\ell_{\rm max}}
\newcommand{\smax}{s_{\rm max}}
\newcommand{\nside}{\ensuremath{n_{\rm side}}}
\newcommand{\npix}{\ensuremath{N_{\rm pix}}}
\newcommand{\Opix}{\ensuremath{\Omega_{\rm pix}}}
\newcommand{\dbar}{\eth}
\newcommand{\bardbar}{\bar{\dbar}}
\newcommand{\dd}{{\rm d}}
\newcommand{\dr}{{\rm dr}}
\newcommand{\bardr}{\bar{\dr}}
\newcommand{\mycode}[1]{\textsc{\tt #1}}
\newcommand{\Planck}{\textit{Planck}}
\newcommand{\Quickbeam}{{\tt Quickbeam}}
\newcommand{\Quickpol}{\mycode{QuickPol}}
\newcommand{\rhwp}{rHWP}

\title{\Quickpol: Fast calculation of effective beam matrices for CMB polarization}
\titlerunning{\Quickpol}
\author{%
Eric Hivon\inst{1},
Sylvain Mottet\inst{1} \&
Nicolas Ponthieu\inst{2,}\inst{3}}
\authorrunning{E. Hivon, S. Mottet \& N. Ponthieu}
\offprints{hivon@iap.fr}
\institute{Sorbonne Universit\'{e}s, UPMC Univ.~Paris 6 \& CNRS (UMR7095): Institut d'Astrophysique de Paris, 98 bis Boulevard Arago, F-75014, Paris, France
\and
Institut de Plan\'etologie et d'Astrophysique de Grenoble,
Universit\'e Grenoble Alpes, CNRS (UMR5274), F-38000, Grenoble, France
\and
Institut d'Astrophysique Spatiale, CNRS (UMR8617) Universit\'e Paris-Sud 11, B\^atiment 121, F-91405, Orsay, France
}
\date{Received Aug 31, 2016 / Accepted Oct 03, 2016}

\begin{document}

\abstract{Current and planned observations of the cosmic microwave background
  (CMB) polarization anisotropies, with their ever increasing number of
  detectors, have reached a potential accuracy that requires a very demanding
  control of systematic effects.  While some of these systematics can be
  reduced in the design of the instruments, others will have to be modeled
  and hopefully accounted for or corrected \emph{a posteriori}. We propose \Quickpol,
    a quick and accurate calculation of the full effective beam transfer
    function and of temperature to polarization leakage at the power spectra
    level, as induced by beam imperfections and mismatches between detector
    optical and electronic responses. All the observation details such as exact
    scanning strategy, imperfect polarization measurements, and flagged samples are accounted for. 
    Our results are validated on \Planck{} high frequency instrument (HFI) simulations. We show how the pipeline
    can be used to propagate instrumental uncertainties up to the final science products, 
    and could be applied to experiments with rotating half-wave plates.
}
\maketitle
 %\tableofcontents

\section{Introduction}
\label{sec:introduction}
We are now entering an era of precise measurements of the cosmic microwave background (CMB) polarization, with
potentially enough sensitivity to detect or even characterize the primordial
tensorial $B$ modes, the smoking gun of inflation
(e.g.,~\cite{ZaldarriagaSeljak1997} and references therein).  This raises
expectations about the control and the correction of contaminations by
astrophysical foregrounds, observational features, and instrumental
imperfections. As it has in the past, progress
will come from the synergy between instrumentation and data
analysis. Improvements in instrumentation call for improved precision in
final results, which are made possible by improved algorithms and the
ability to deal with more and more massive data sets. In turn,
expertise gained in data processing allows for better simulations that lead to new
instrument designs and better suited observations. An example of such joint
developments is the study of the impact of optics- and electronics-related
imperfections on the measured CMB temperature and polarization angular power
spectra and their statistical isotropy. Systematic effects such as beam non-circularity, 
response mismatch in dual polarization measurements and scanning strategy imperfections, 
as well as how they can be mitigated, have been extensively studied
in the preparation of forthcoming instruments \citep[including, but not limited to][]{%
Souradeep+2001, Fosalba+2002, Hu+2003, Mitra+2004,
Mitra+2009, ODea+2007, Rosset+2007, Shimon+2008, Miller+2009a, Miller+2009b,
Hanson+2010, Leahy+2010, Rosset+2010, Ramamonjisoa+2013, Rathaus+2014,
Wallis+2014, Pant+2015},
and during the analysis of data collected by
WMAP\footnote{Wilkinson microwave anisotropy probe: \url{http://map.gfsc.nasa.gov}.} \citep{Smith+2007, Hinshaw+2007,
    Page+2007} or \Planck\footnote{\url{http://www.esa.int/Planck}.}
  \citep{Planck2013-7, Planck2013-17, Planck2015-11} satellite
  missions. 
\\
At the same time, several deconvolution algorithms and codes have been proposed
to clean up the CMB maps from such beam-related effects prior to the computation
of the power spectra, like \mycode{PreBeam} \citep{Prebeam2009},
\mycode{ArtDeco} \citep{Artdeco2012}, and in \citet{Bennett+2013} and
\citet{Wallis+2015}.
\\
Finally, in a related effort, the \mycode{FEBeCoP} pipeline, 
described in \citet{Mitra+2011} and used in
\Planck{} data analysis \citep{Planck2013-4, Planck2013-7}, 
can be seen as a convolution facility, by providing, 
at arbitrary locations on the sky, the effective beam maps and 
point spread functions of a detector set, which,
in turn, can be used for a Monte-Carlo based description 
of the effective beam window functions for a given sky model.

In this paper, we introduce the \Quickpol{} pipeline, an extension to
polarization of the \Quickbeam{} algorithm used in
\citet{Planck2013-7}. It allows a quick and accurate computation of the
leakage and cross-talk between the various temperature and polarization power
spectra ($TT$, $EE$, $BB$, $TE$, etc.) taking into account the exact scanning,
sample flags, relative weights, and scanning beams of the considered
set(s) of detectors.  The end results are effective beam matrices describing, for
each multipole $\ell$ , the mixing of the various spectra, independently of the
actual value of the spectra.  As we shall see, the impact of changing any
time-independent feature of the instrument, such as its beam maps, relative gain
calibrations, detector orientations, and polarization efficiencies can be
propagated within seconds to the final beam matrices products, allowing
extremely fast Monte-Carlo exploration of the experimental
features. \Quickpol{} is thus a powerful tool for both real data analysis
  and forthcoming experiments, simulations and design.

The paper is organized as follows. The mathematical formalism is exposed in Section~\ref{sec:formalism} 
and analytical results are given in Section~\ref{sec:results}.
The numerical implementation is detailed in Section~\ref{sec:numerical_implementation}
and compared to the results of \Planck{} simulations in Section~\ref{sec:simulations}.
Section~\ref{sec:propagation} shows the propagation of instrumental uncertainties. We discuss briefly the case of rotating half-wave plates in
Section~\ref{sec:rhwp} and conclude in
Section~\ref{sec:conclusions}.

\section{Formalism}
\label{sec:formalism}
\subsection{Data stream of a polarized detector}
\label{sec:datastream}

As usual in the study of polarization measurement, we will use
Jones' formalism to study the evolution of the electric
component of an electro-magnetic radiation in the optical system.  Let us
consider a quasi monochromatic\footnote{Although it is important when trying
  to disentangle sky signals with different electromagnetic spectra
  \citep{Planck2013-6}, the finite bandwidth of the actual detectors only plays
  a minor role in the problem considered here, and will be ignored in this paper.} radiation
propagating along the $z$ axis, and hitting the optical system at a position
$\vecr = \vectortwo{x}{y}$. The incoming electric field $\vece(\vecr) =
\vectortwo{e_x}{e_y} e^{i k(z - c t)}$ will be turned into $\vece'(\vecr) =
\matJ(\vecr) . \vece(\vecr)$, where $\matJ(\vecr)$ is the 2x2 complex Jones
matrix of the system.

A rotation of the optical system by $\alpha$ around the $z$ axis can be seen as
a rotation of both the orientation and location of the incoming radiation by
$-\alpha$ in the detector reference frame, and the same input radiation
is now received as
\begin{equation}
        \vece'(\alpha, \vecr) = \matJ\left(\vecr_\alpha\right) \mydot \matR^\dagger_\alpha \mydot \vece(\vecr),
\end{equation}
with 
\begin{align}
\vecr_\alpha &= \matR^\dagger_\alpha . \vecr, \\
\matR_\alpha &= \jm{\cos\alpha}{-\sin\alpha}{\sin\alpha}{\cos\alpha},
\end{align}
and the $\dagger$ sign representing the adjoint operation, which for a
real rotation matrix, simply amounts to the matrix transpose.
The measured signal is
\begin{equation}
        d(\alpha) = \int {\rm d}\vecr \ d(\alpha, \vecr)
\end{equation}
with
\begin{align}
d(\alpha, \vecr) & = \VEV{\vece'^\dagger . \vece'} = \VEV{\trace \left(\vece'.\vece'^\dagger \right) }
\nonumber \\
        & = \trace\left( 
\matJ(\vecr_\alpha) \mydot \matR^\dagger_\alpha 
                   \mydot \VEV{\vece.\vece^\dagger} 
                   \mydot \matR_\alpha 
                   \mydot \matJ^\dagger(\vecr_\alpha) \right).
\end{align}
We now introduce the Stokes parameters of the input signal (dropping the dependence on $\vecr$)
\begin{equation}
   \VEV{\vece . \vece^\dagger} = \frac{1}{2}\jm{T+Q}{U+iV}{U-iV}{T-Q}
\end{equation}
and of the (un-rotated) instrument response
\begin{equation}
\matJ^\dagger.\matJ = 
\frac{1}{2}\jm
{\bI +\bQ}%
{\bU-i\bV}%
{\bU+i\bV}%
{\bI -\bQ},%
\end{equation} to obtain
\begin{align}
        d(\alpha) &= \frac{1}{2} \int {\rm d}\vecr 
        \left[ 
         \bI (\alpha, \vecr)T(\vecr)
        +\bQ (\alpha, \vecr)Q(\vecr)
        +\bU (\alpha, \vecr)U(\vecr)
\onetwocols{\cols}{}{\right. \nonumber \\ &\quad\quad \left.}
        -\bV (\alpha, \vecr)V(\vecr) \right].
        \label{eq:datastream_beam}
\end{align}
With the rotated instrument response:
\begin{subequations}
\label{eq:beamstokesrotation}
\begin{align}
\bI (\alpha, \vecr) &= \bI(\vecr_\alpha), \\
\bQ (\alpha, \vecr) &= \bQ(\vecr_\alpha)\cos 2\alpha - \bU(\vecr_\alpha)\sin 2\alpha, \\
\bU (\alpha, \vecr) &= \bQ(\vecr_\alpha)\sin 2\alpha + \bU(\vecr_\alpha)\cos 2\alpha, \\
\bV (\alpha, \vecr) &= \bV(\vecr_\alpha).
\end{align}
\end{subequations}
Following \cite{Rosset+2010}, we can specify the instrument as being a beam forming optics, followed by 
an imperfect polarimeter in the direction $x$, with $0\le \eta \le 1$, and having an overall optical efficiency $0\le\tau\le 1$:
\begin{equation}
        \matJ(\vecr) = 
        \sqrt{\tau}
        \jm{1}{0}{0}{\sqrt{\eta}}
        \jm{b_{xx}(\vecr)}{b_{xy}(\vecr)}{b_{yx}(\vecr)}{b_{yy}(\vecr)},
\end{equation}
with 
\begin{equation}
        \vectortwo{b^*_{ax}}{b^*_{ay}}.\left(b_{ax}\ b_{ay}\right) = 
        \frac{1}{2}\jm{\bI_a+\bQ_a}{\bU_a-i\bV_a}{\bU_a+i\bV_a}{\bI_a-\bQ_a}
\end{equation}
for $a = x,y$. The Stokes parameters of the instrument are then
$\bS = \tau (\bS_x + \eta \bS_y)$ for $\bS=\bI, \bQ, \bU, \bV$.

If the beam is assumed to be perfectly co-polarized, that is, it does not alter at all the polarization of the incoming radiation,
with
$b_{xy}=b_{yx}=0$ and $b_{xx}=b_{yy}$, then $\bU_x=\bU_y=\bV_x=\bV_y=0$, $\bI_x=\bI_y=\bQ_x=-\bQ_y$, and
$\bI = (1+\eta)\bI_x$, 
$\bQ = (1-\eta)\bQ_x$, 
$\bU=\bV=0$, Eqs.~(\ref{eq:datastream_beam}, \ref{eq:beamstokesrotation}) become
\begin{equation}
        d(\alpha) = \frac{1+\eta}{2} \tau \int {\rm d}\vecr 
        \bI_x(\vecr_\alpha) \left[
        T(\vecr) + \cpe\left(Q(\vecr)\cos2\alpha + U(\vecr)\sin2\alpha\right)
        \right],
        \label{eq:datastream_copolar}
\end{equation}
where
\begin{equation}
  \cpe = \frac{1-\eta}{1+\eta}
\end{equation}
is the polar efficiency, such that $0 \le \cpe \le 1$ with $\cpe =1$ for a
perfect polarimeter and $\cpe=0$ for a detector only sensitive to intensity. In
the case of \Planck{} high frequency instrument (HFI), \citet{Rosset+2010} showed the measured polarization
efficiencies to differ by $\Delta \cpe' = $1\% to 16\% from their ideal values,
with an absolute statistical uncertainty generally below 1\%.
The particular case of co-polarized beams is important because in most experimental setups, 
such as \Planck, the beam response
calibration is done on astronomical or artificial far field sources. Well known,
compact, and polarized sources are generally not available to measure $\bQ$ and $\bU$
and only the intensity beam response $\bI$ is measured. In the absence of
reliable physical optics modeling of the beam response, one therefore has to
assume $\bQ$ and $\bU$ to be perfectly co-polarized.

So far, we have only considered the optical beam response.  We should also
take into account the scanning beam, which is the convolution of the optical beam with
the finite time response of the instrument (or its imperfect correction) as it
moves around the sky, as described in \citet{Planck2013-7} and
\citet{Planck2015-7}. These time related effects can be a major source of
elongation of the scanning beams, and can increase the beam mismatch among
sibling detectors. If one assumes the motion of the detectors on the sky to be
nearly uniform, as was the case for \Planck, then optical beams can readily be
replaced by scanning beams in the \Quickpol{} formalism.

\subsection{Spherical harmonics analysis}
\label{sec:SH}
We now define the tools that are required to extend the above results to
  the full celestial sphere.  The temperature $T$ is a scalar
quantity, while the linear polarization $Q \pm i U$ is of spin $\pm 2$, and
the circular polarization $V$ is generally assumed to vanish.  They can be written as linear combinations of
spherical harmonics (SH):
\begin{align}
        T(\vecr)               &= \sum_{\ell m}        a^T_{\ell m}\           Y_{\ell m}(\vecr),
\label{eq:Tlm} \\
        Q(\vecr)\pm i U(\vecr) &= \sum_{\ell m} {}_{\pm2}a_{\ell m}\  {}_{\pm2}Y_{\ell m}(\vecr),
\label{eq:QUlm}
\end{align}
although one usually prefers the scalar and fixed parity $E$ and $B$ components
\begin{equation}
   a^{E}_{\ell m} \pm i a^{B}_{\ell m} =        -\ {}_{\pm 2}a_{\ell m}
\label{eq:spin_stokes}
\end{equation}
such that $a^{X*}_{\ell m} = (-1)^m a^{X}_{\ell -m}$ for $X=T,E,B$.
In other terms
\begin{equation}
        \vectorthree{\,_{0} a_{\ell m}}{\,_{2}a_{\ell m}}{\,_{-2}a_{\ell m}} 
        = \matR_2. \vectorthree{a^T_{\ell m}}{a^E_{\ell m}}{a^B_{\ell m}}
        \label{eq:alm_spin_stokes}
\end{equation}
with
\begin{equation}
        \matR_2      = \matrixthree{1}{0}{0}{0}{ -1}{ -i}{0}{  -1}{  i}.
        \label{eq:rotation_matrix2_def_main}
\end{equation}
The sign convention used in Eq.~(\ref{eq:spin_stokes})
is consistent with \citet{ZaldarriagaSeljak1997} and the HEALPix\footnote{\url{http://healpix.sourceforge.net}.} library \citep{Gorski+2005}.

The response of a beam centered on the North pole can also be decomposed in SH coefficients
\begin{align}
        b_{\ell m} &= \int \dd\vecr \bI(\vecr) Y^*_{\ell m}(\vecr),
\label{eq:bTlm} \\
        _{\pm 2} b_{\ell m} &= \int \dd\vecr \left(\bQ(\vecr)\pm i\bU(\vecr)\right)\ {}_{\pm2}Y^*_{\ell m}(\vecr),
\label{eq:bQUlm}
\end{align}
while the coefficients of a rotated beam can be computed by noting that
under a rotation of angle $\alpha$ around the direction $\vecr$, the SH of spin $s$ transform as
\begin{align}
        _s Y_{\ell m}(\vecr')&\longrightarrow \sum_{m'}\ _s Y_{\ell m'}(\vecr') D^{\ell}_{m'm}(\vecr,\alpha)
        \label{eq:rotYlm}.
\end{align}
The elements of Wigner rotation matrices $D$ are related to the SH via \citep{Challinor+2000}
\begin{align}
        D^{\ell}_{m'm}(\vecr,\alpha) &= (-1)^m q_{\ell}\ _{-m} Y_{\ell m'}^*(\vecr)e^{-i m \alpha},
        \label{eq:defWignerRot}
\end{align}
with $q_{\ell} = \sqrt{\frac{4\pi}{2\ell+1}}$. 

If the beam is assumed to be
co-polarized and coupled with a perfect polarimeter rotated by an angle $\gamma$,
such that $\bQ+i\bU = \bI e^{2 i \gamma}$ in cartesian coordinates (or $\bQ+i\bU
= \bI e^{2 i (\gamma-\phi)}$ in $(\theta, \phi)$ polar coordinates), simple relations between 
$b_{\ell m}$ and $_{\pm 2} b_{\ell,m}$ can be established.
For a Gaussian circular beam of full width half maximum (FWHM) $\fwhm=\sigma
\sqrt{8 \ln 2}\approx 2.355 \sigma$ and of throughput $\int \dd\vecr \bI(\vecr)
= \sqrt{4\pi}\ b_{00} = 1,$
\citet{Challinor+2000} found
\begin{subequations}
\label{eq:gaussbeam}
\begin{align}
        b_{\ell m} &= \sqrt{\frac{2\ell+1}{4\pi}}e^{ -\frac{1}{2}\ell(\ell+1)\sigma^2}
        \ \delta_{m,0}
              \label{eq:gaussbeam_T},
\\
        _{\pm 2} b_{\ell,m} &= b_{\ell,m \pm 2}\ e^{2\sigma^2}\ e^{\pm 2i\gamma}.
                \label{eq:gaussbeam_P}
\end{align}
\end{subequations}
The factor $c_2=e^{2\sigma^2}$ in Eq.~(\ref{eq:gaussbeam_P}) is such that
$c_2-1 < 1.1\ 10^{-4}$ for $\fwhm \le 1\degr$ and $c_2-1 < 3.1\ 10^{-6}$ for
$\fwhm\le 10\arcmin$, and will be assumed to be $c_2=1$ from now on.
For a slightly elliptical Gaussian beam, \citet{Fosalba+2002} found
\begin{equation}
        _{\pm 2} b_{\ell,m} = b_{\ell, m\pm 2}\  e^{\pm 2i\gamma},
        \label{eq:egaussbeam_P}
%\end{align}
\end{equation}
while we show in Appendix \ref{sec:copolar_beams} that Eq.~(\ref{eq:egaussbeam_P}) is true for arbitrarily shaped co-polarized beams.
This result can also be obtained by noting that an arbitrary beam is the sum of Gaussian circular beams 
with different FWHM and center \citep{Asymfast2004}, each of them obeying Eq.~(\ref{eq:gaussbeam_P}).

The detector
associated to a beam is an imperfect polarimeter with a polarization
efficiency $\cpe'$ and the overall polarized response of the detector,
in a referential \emph{aligned} with its direction of polarization (the so-called Pxx
coordinates in \Planck{} parlance), reads
\begin{equation}
        \bQ = \cpe'\bI,
\end{equation}
so that
\begin{equation}
        _{\pm 2} b_{\ell,m} = \cpe' b_{\ell,m \pm 2}.
   % e^{\pm 2i\gamma}.
\end{equation}
We introduced $\cpe'$ to distinguish it from the $\cpe$ value used in the map-making, as described below.

%%-----------------
\subsection{Map making equation}

A polarized detector pointing, at time $t$, in the direction $\vecr_t$ on the sky, 
and being sensitive to the polarization with angle $\alpha_t$ with respect to the local meridian, measures
\begin{align}
        d(\vecr_t,\alpha_t) &= \int \dd\vecr' \left[ \bI(\vecr_t,\alpha_t; \vecr') T(\vecr')
           + \bQ(\vecr_t,\alpha_t; \vecr') Q(\vecr')  
\onetwocols{\cols}{}{\right. \nonumber \\  &\quad\quad \left.}
           + \bU(\vecr_t,\alpha_t; \vecr') U(\vecr') \right].
        \label{eq:TODanybeam}
\end{align}
The factor $1/2$ present in Eq.~(\ref{eq:datastream_beam}) is assumed to be absorbed in the gain calibration, performed on large scale temperature fluctuations, such as the CMB solar dipole \citep{Planck2013-8}, and we assumed the circular polarization $V$ to vanish. With the definitions introduced in Section \ref{sec:SH}, this becomes
\begin{align}
d(\vecr_t,\alpha_t)     &= \sum_{\ell ms} \left[ 
                          _{0} a_{\ell m}\ _{0} b^{*}_{\ell s} + 
                1/2\left( _{2} a_{\ell m}\ _{2} b^{*}_{\ell s} +
                        \ _{-2} a_{\ell m}\ _{-2} b^{*}_{\ell s} \right)\right]
\onetwocols{\cols}{}{\nonumber\\ &\quad\quad \times}
  (-1)^s\, q_{\ell}\, e^{is\alpha_t} \ _{-s}Y_{\ell m}(\vecr_t).
        \label{eq:TOD_from_alm}
\end{align}

The map-making formalism is set ignoring the beam effects, assuming a perfectly
co-polarized detector and an instrumental noise $n$ \citep[][and references therein]{Polkapix2011}, 
so that, for a detector $j$, Eq.~(\ref{eq:datastream_copolar}) becomes
\begin{align}
        d_j(t) &= T(p) + \cpem{j} Q(p) \cos 2\alpha^{(j)}_t + \cpem{j} U(p) \sin 2\alpha^{(j)}_t + n_j(t),
        \label{eq:ideal_tod}
\end{align}
where the leading prefactors are here again absorbed in the gain
calibration. Let us rewrite it as
\begin{align}
          d_j(t) &= A^{(j)}_{t,p} m(p) + n_j(t),
        \label{eq:ideal_tod_matrix}
\end{align}
with \citep{Shimon+2008} 
\begin{align}
A^{(j)}_{t,p} &= 
\left(1, \cpem{j}e^{-2i\alpha^{(j)}_t}, \cpem{j}e^{2i\alpha^{(j)}_t}\right),  \label{eq:Amatrix} \\
        m(p) &= \left(T, P/2, P^*/2\right)^T,
\end{align}
and $P = Q+iU$. Assuming the noise to be uncorrelated between detectors, with
covariance matrix $\matN_j = \VEV{\vecn_j \mydot \vecn_j^T}$ for detector $j$, the generalized
least square solution of Eq.~(\ref{eq:ideal_tod}) for a set of detectors is
\begin{equation}
        \tvm = \left(\sum_k \matA^{(k)\dagger} \mydot \matN_k^{-1} \mydot \matA^{(k)} \right)^{-1} 
      \mydot \sum_j \matA^{(j)\dagger} \mydot \matN_j^{-1} \mydot \vecd_j.
        \label{eq:map_making_top}
\end{equation}
Let us now replace the ideal data stream (Eq.~\ref{eq:ideal_tod}) with the one
obtained for arbitrary beams (Eq.~\ref{eq:TODanybeam}) and further
  assume that the
noise is white and stationary with variance $\sigma_{j}^2$, so that
$\matN_j^{-1} = 1/\sigma_{j}^2 = w_j$. Let us also introduce the binary flag $f_{j,t}$
used to reject individual time samples from the map-making process;
Eq.~(\ref{eq:map_making_top}) then becomes
\begin{align}%---------
\tvm(p) &\equiv \vectorthree{\tm(0;p)}{\tm(2;p)/2}{\tm(-2;p)/2}, \\
         & = \left(\sum_k \sum_{t\in p} A^{(k)\dagger}_{p,t} w_k f_{k,t} A^{(k)}_{t,p} \right)^{-1} 
                 \left(\sum_j \sum_{t\in p} A^{(j)\dagger}_{p,t} w_j f_{j,t}
                 d_{j,t} \right ).
\label{eq:tvmp}
\end{align}%---------                                                                                                                        
We have assumed here the pixels to be infinitely small, so that, starting with Eq.~(\ref{eq:TOD_from_alm}), 
the location of all samples in a pixel coincides 
with the pixel center. The effect of the pixel's finite size and the so-called sub-pixel effects 
will be considered in Section \ref{sec:subpixel}.

\subsection{Measured power spectra}
\label{sec:measured_cl}
To compute the cross-power spectrum of any two spin $v_1$ and $v_2$ maps,
we first project each polarized component $v$ of $\tvm(p)$ on the appropriate spin weighted sets of
spherical harmonics,
\begin{equation}
_x \tm_{\ell''m''}(v) = \int \dd\vecr\  \tm(v;\vecr)\ _{x}Y^*_{\ell''m''}(\vecr),
\label{eq:proj_sph}
\end{equation}
and average these terms according to
\begin{align}
 \pC^{v_1v_2}_{\ell''} &\equiv
      \frac{1}{2\ell''+1}\sum_{m''} 
        \VEV{\ _{v_1} \tm_{\ell''m''}(v_1)  
        \ _{v_2} \tm^*_{\ell''m''}(v_2)}, \\
        &= \sum_{u_1u_2 j_1 j_2 \ell s_1s_2} %e^{i(v_1\gamma_{j_1}-v_2\gamma_{j_2})}
        (-1)^{s_1+s_2+v_1+v_2} \, C^{u_1u_2}_{\ell}
        \frac{2\ell+1}{4\pi} 
        \ _{u_1} \hatb^{(j_1)*}_{\ell s_1} \  _{u_2} \hatb^{(j_2)}_{\ell s_2} 
 \nonumber \\ &\quad \times
        \frac{k_{u_1}k_{u_2}}{k_{v_1}k_{v_2}} 
        \sum_{\ell'm'}   \cpem{j_1,v_1}\ \cpem{j_2,v_2}
        \ {}_{s_1+v_1}\tomega^{(j_1)}_{\ell'm'}
        \ {}_{s_2+v_2}\tomega^{(j_2)*}_{\ell'm'}
\onetwocols{\cols}{}{\nonumber \\ &\quad \times}
        \wjjj{\ell}{\ell'}{\ell''}{-s_1}{s_1+v_1}{-v_1}
        \wjjj{\ell}{\ell'}{\ell''}{-s_2}{s_2+v_2}{-v_2}
        ,
        \label{eq:crossCl_general}
\end{align}
where Eq.~(\ref{eq:w3j_ortho2}) was used. The detailed derivation of this
  relation and its associated terms is given in Appendix \ref{appendix:projection_sh}. Suffice it to say
  here that $k_u$ terms are either $1$ or $1/2$, $_{u} \hat{b}_{\ell s}^{(j)}$ terms are inverse
  noise-weighted beam multipoles, and $\tomega^{(j)}$ terms are effective weights describing the scanning and
  depending on  
  the direction of polarization, hit redundancy
  (both from sky coverage and flagged samples), and noise level of detector $j$.

  Equation~(\ref{eq:crossCl_general}) is therefore a generalization to
non-circular beams of the pseudo-power spectra measured on a masked or weighted
map \citep{Hivon+2002, Hansen+2003}, and extends to polarization the \Quickbeam{} 
non-circular beam formalism used in the data analysis conducted by \citet{Planck2013-7}.
 It also formally agrees with
\citet{Hu+2003}'s results on the impact of systematic effects on the polarization
power spectra, with the functions $\ _{u} \hatb^{(j)*}_{\ell s}
\cpem{j,v}\ {}_{s+v}\tomega^{(j)}_{\ell'm'}$ absorbing the systematic effect
  parameters relative to detector $j$. In the next sections, we present
  the numerical results implied by this result and compare them on full-fledged
  Planck-HFI simulations.

\section{Results}
\label{sec:results}
We now apply the \Quickpol{} formalism to configurations representative of current or forthcoming CMB experiments,
and to a couple of idealized test cases for which the expected result is already known, as a sanity check. 
The effect of the finite pixel size is also studied.

%-----------------------------------------------------------------
\begin{figure*}[htbp]
%-----------------------------------------------------------------
   \includegraphics[angle=90,width=0.35\textwidth,trim={30pt 110 130 0},clip]{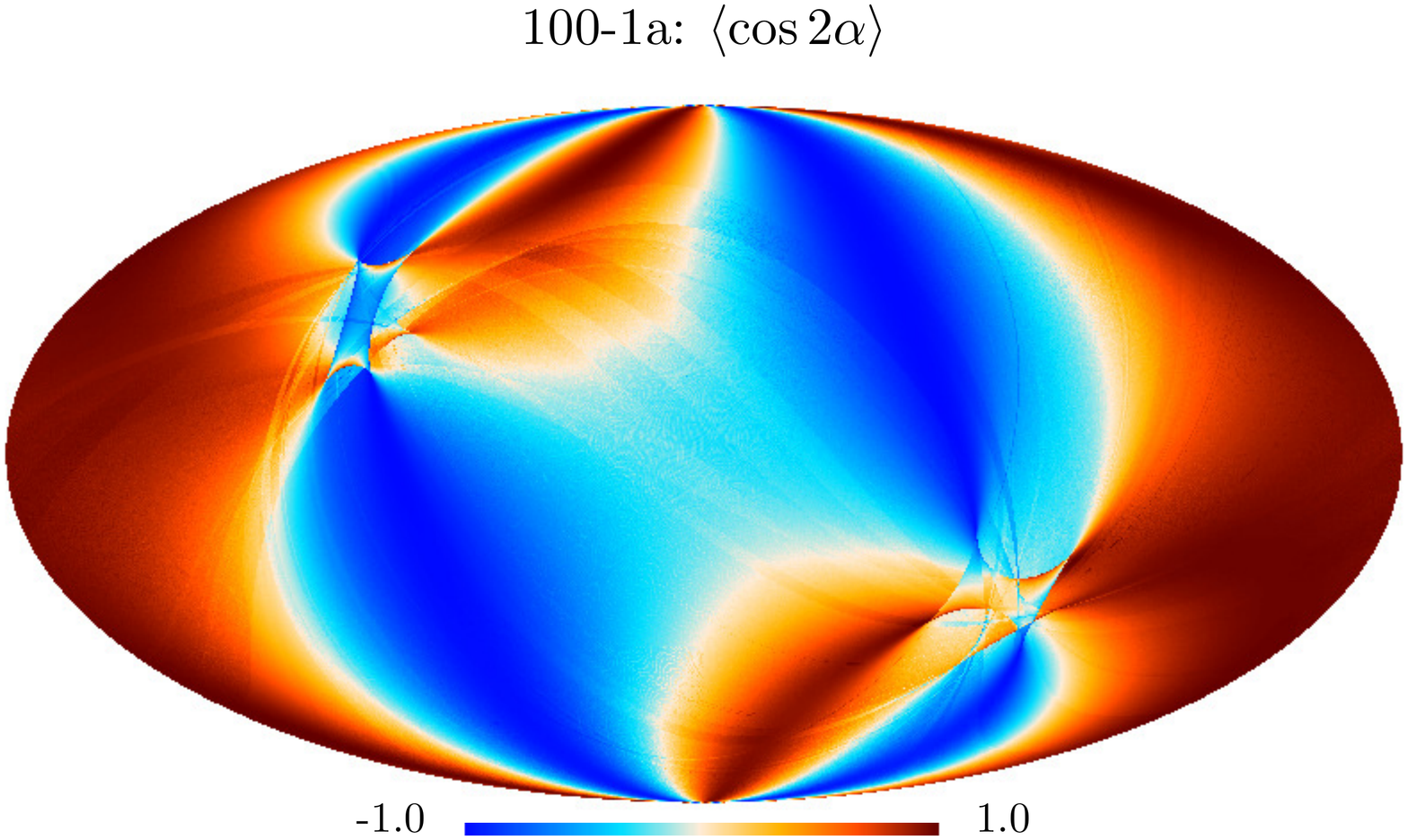}
   \includegraphics[angle=90,width=0.35\textwidth,trim={30pt 110 130 0},clip]{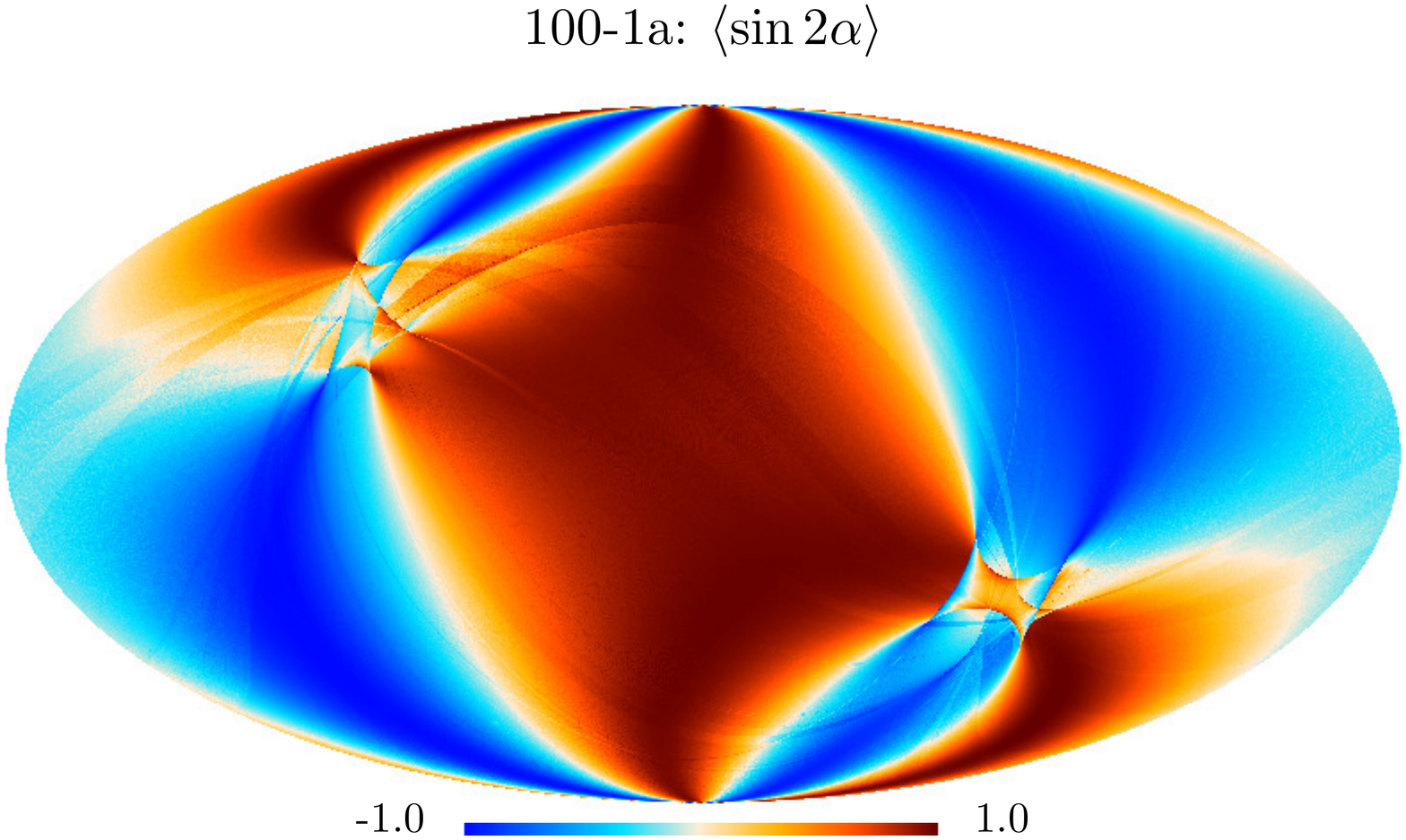}
   \includegraphics[angle=90,width=0.29\textwidth,trim={0pt    0   0 0},clip]{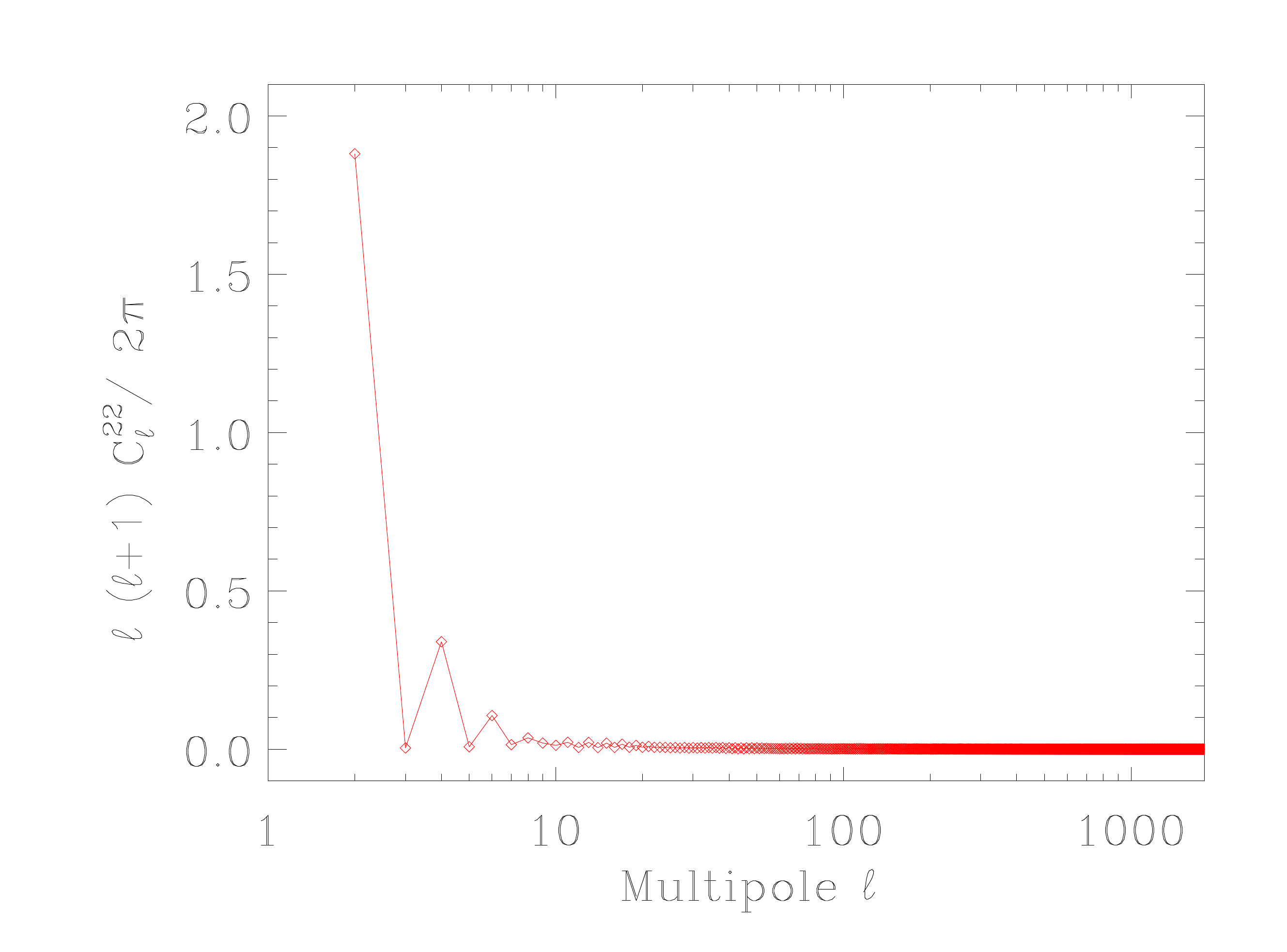}
  \caption{\label{fig:hits_Planck} Orientation of polarization measurements in
    \Planck. The two left panels show, for an actual \Planck{} detector, the
    maps of $\VEV{\cos 2\alpha}$ and $\VEV{\sin 2\alpha}$ respectively, where
    $\alpha$ is the direction of the polarizer with respect to the local
    Galactic meridian, which contributes to the spin 2 term $\omega_2^{(j)}$
    defined in Eq.~(\ref{eq:spin_omega}).  The right panel shows the power
    spectrum $C^{22}_{\ell}$ of
    $\VEV{e^{2i\alpha}}=\omega_2^{(j)}/\omega_0^{(j)}$, multiplied by
    $\ell(\ell+1)/2\pi$.}
\end{figure*}
%-----------------------------------------------------------------

%-----------------------------------------------------------------
\begin{figure*}[htbp]
%-----------------------------------------------------------------
%    \includegraphics[angle=90,width=0.35\textwidth,trim={30pt   0   0 0},clip]{fig_quickpol/hit_LC45_c2.pdf}
%    \includegraphics[angle=90,width=0.35\textwidth,trim={30pt   0   0 0},clip]{fig_quickpol/hit_LC45_s2.pdf}
%    \includegraphics[angle=90,width=0.29\textwidth,trim={0pt    0   0 0},clip]{fig_quickpol/cl_LC45_22.pdf}
%    \includegraphics[angle=90,width=0.35\textwidth,trim={30pt   0   0 0},clip]{fig_quickpol/hit_LB45_c2.pdf}
%    \includegraphics[angle=90,width=0.35\textwidth,trim={30pt   0   0 0},clip]{fig_quickpol/hit_LB45_s2.pdf}
%    \includegraphics[angle=90,width=0.29\textwidth,trim={0pt    0   0 0},clip]{fig_quickpol/cl_LB45_22.pdf}
   \includegraphics[angle=90,width=0.35\textwidth,trim={30pt   0   0 0},clip]{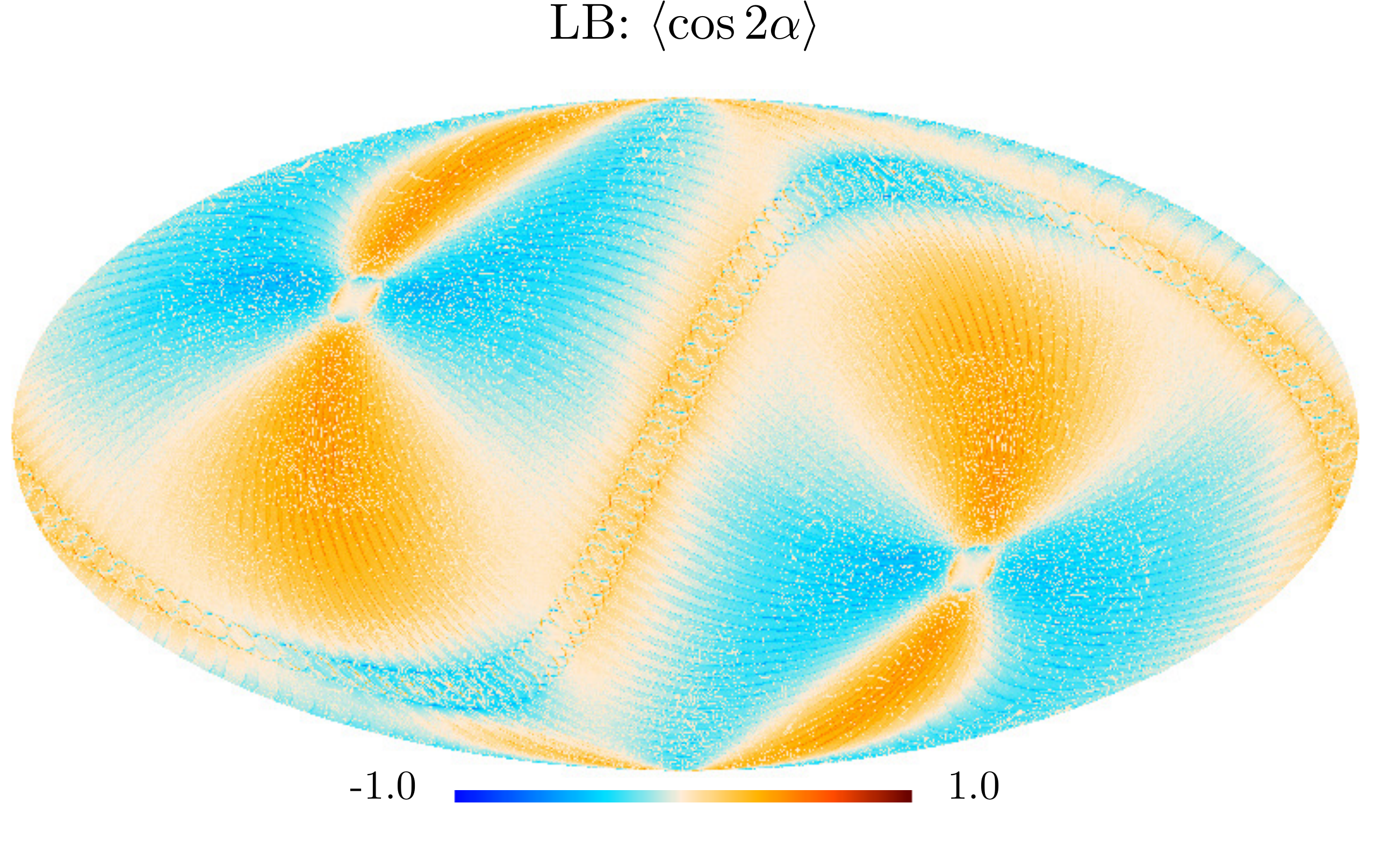}
   \includegraphics[angle=90,width=0.35\textwidth,trim={30pt   0   0 0},clip]{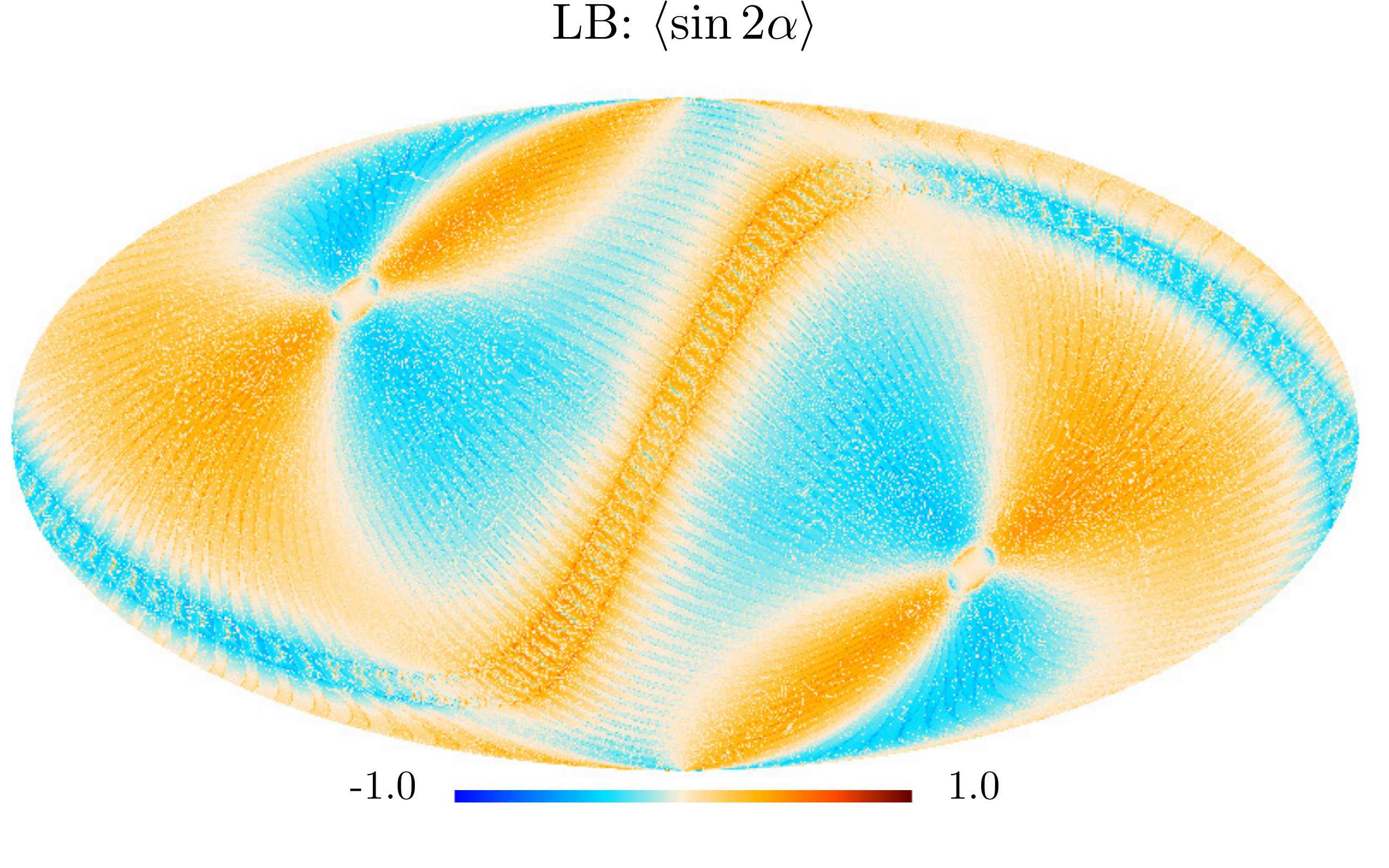}
   \includegraphics[angle=90,width=0.29\textwidth,trim={0pt    0   0 0},clip]{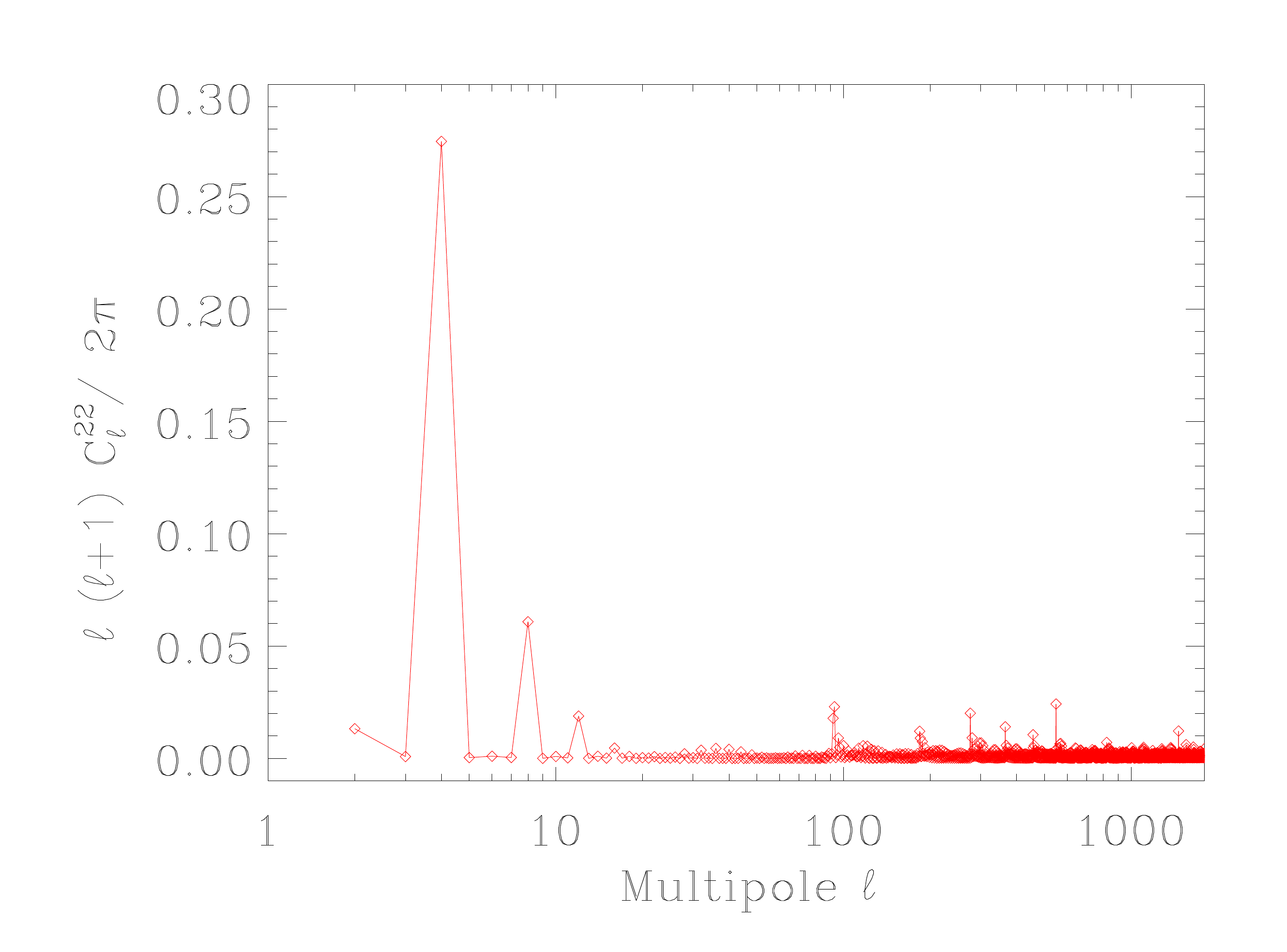}
  \caption{\label{fig:hits_LB} Same as Fig.~\ref{fig:hits_Planck} for an hypothetical detector of a LiteBIRD-like mission,
except for the right panel plot which has a different $y$-range.}
\end{figure*}
%-----------------------------------------------------------------

%-----------------------------------------------------------------
\begin{figure*}[htbp]
%-----------------------------------------------------------------
  \includegraphics[angle=0,width=0.49\textwidth,trim={10pt 10 50 10},clip]{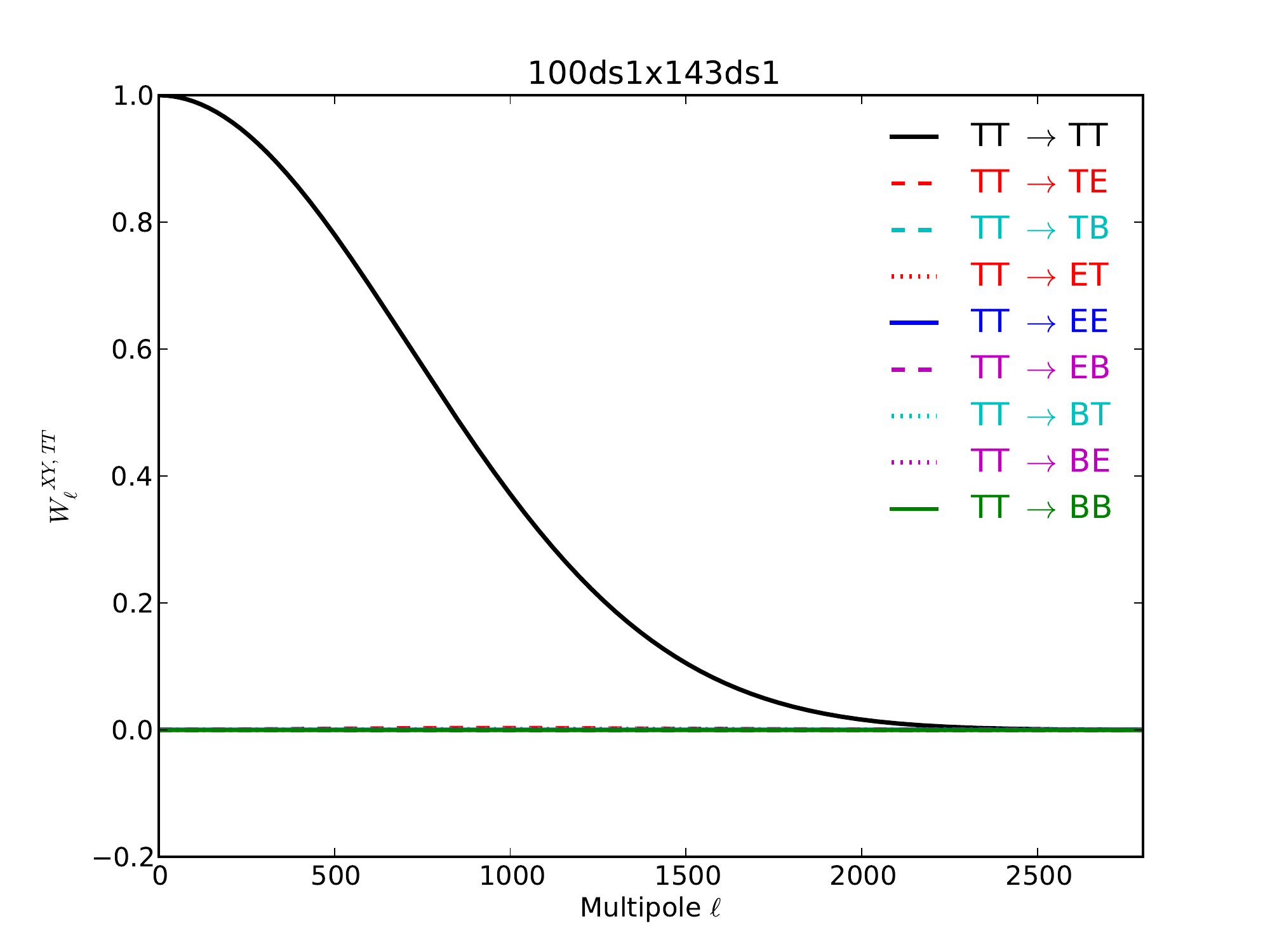}
  \includegraphics[angle=0,width=0.49\textwidth,trim={10pt 10 50 10},clip]{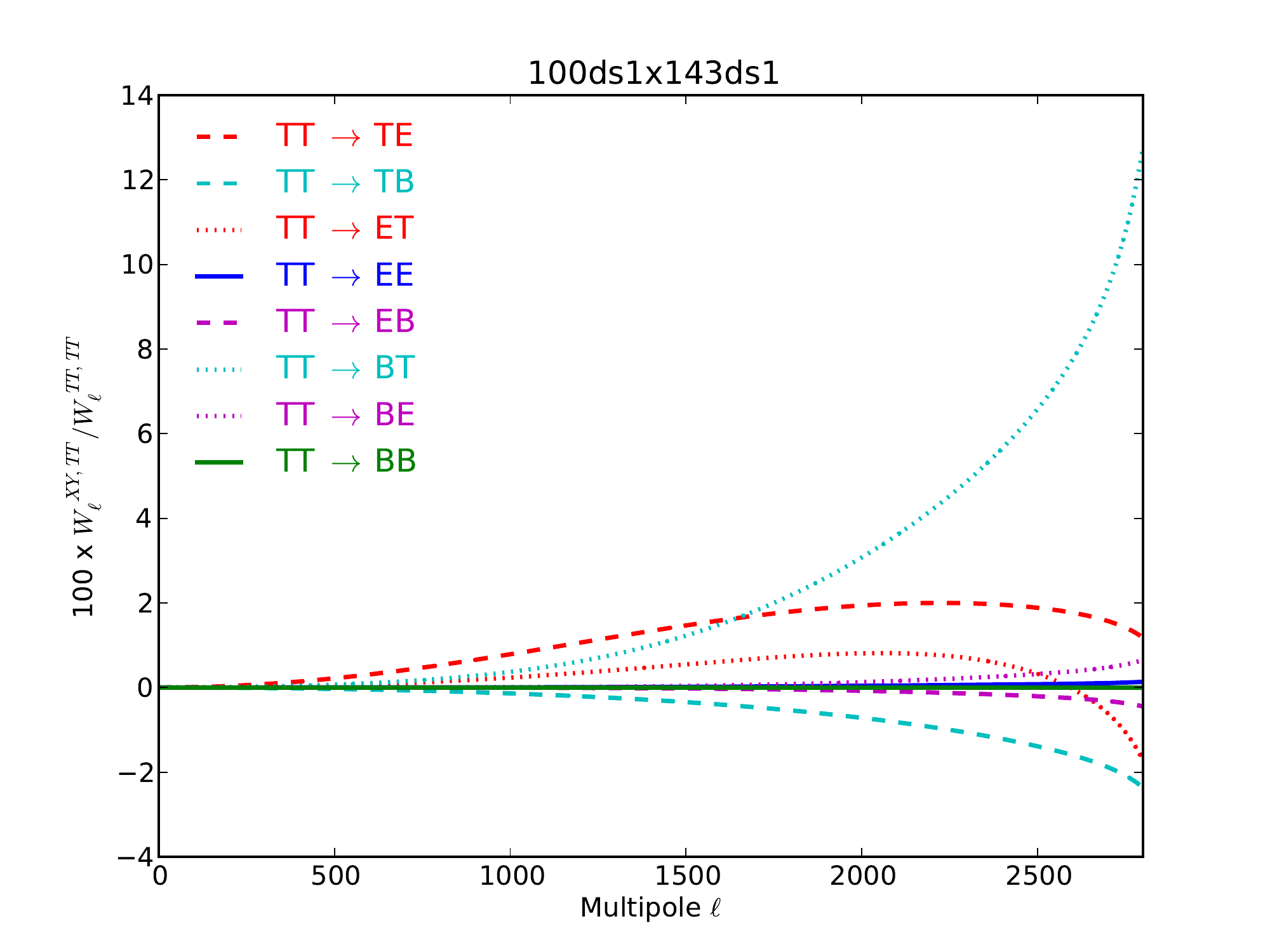}
  \caption{\label{fig:Wmatrix} Effective beam window
    matrix $W_{\ell}^{XY,\; TT}$ introduced in Eq.~(\ref{eq:beam_matrix}) and detailed in Eq.~(\ref{eq:wl_xyTT}), 
for the cross-spectra of two simulated \Planck{} maps
discussed in Section~\ref{sec:simulations}. 
Left panel: raw elements of
$W_{\ell}^{XY,\; TT}$, showing for each $\ell$ how the measured $XY$ map angular
power spectrum is impacted by the input $TT$ spectrum, because of the observation of the sky 
with the beams. 
Right panel: blown-up ratio of the non-diagonal elements to the diagonal
ones: $100\ W_{\ell}^{XY,\; TT}/W_{\ell}^{TT,\; TT.}$}
\end{figure*}
%-----------------------------------------------------------------

\subsection{A note about scanning strategies}
To begin with, let us
consider the scanning strategy of \Planck{} and of another satellite mission
optimized for the measurement of CMB polarization.\\
Figure \ref{fig:hits_Planck} illustrates the orientation of the polarization
measurements achieved in \Planck.  It shows, for an actual \Planck{} detector,
the maps of $\VEV{\cos 2\alpha}$ and $\VEV{\sin 2\alpha}$ respectively, where
$\alpha$ is the direction of the polarizer with respect to the local Galactic
meridian. These quantities contribute to the spin 2 term $\omega_2^{(j)}$
defined in Eq.~(\ref{eq:spin_omega}). The large amplitude of these two maps is
consistent with the fact that for a given detector, the orientation of the
polarization measurements is mostly $\alpha$ and $-\alpha$, as expected when
detectors move on almost great circles with very little precession.  Another
striking feature is the relative smoothness of the maps, which translate into
the power spectrum $C^{22}_{\ell}$ of
$\VEV{e^{2i\alpha}}=\omega_2^{(j)}/\omega_0^{(j)}$ peaking at low $\ell$ values.

Figure \ref{fig:hits_LB} shows the same information for an hypothetical
LiteBIRD\footnote{\url{http://litebird.jp/eng/}.} like detector (but 
\emph{without} half-wave plate modulation) in which we assumed the detector to
cover a circle of 45$\degr$ in radius in one minute, with its spin axis precessing 
with a period of
four days at 50$\degr$ from the anti-sun direction.  As expected for such a
scanning strategy, the values of $\alpha$ are pretty uniformly distributed over
the range $[0, 2\pi]$, which translates into a low amplitude of the $\VEV{\cos
  2\alpha}$ and $\VEV{\sin 2\alpha}$ maps. Even if those maps do not look as smooth
as those of \Planck, their power spectra peak at fairly low multipole
values.

% %%%%%%%%%%%%% General case %%%%%%%%
\subsection{Arbitrary beams, smooth scanning case}
If one assumes that ${\omega}_s(p)$ and ${\tomega}_s(p)$  vary
slowly across the sky, as we just saw in the case of \Planck{} and
LiteBIRD - and probably a wider class of orbital and sub-orbital missions - then
$_s {\tomega}_{\ell'm'}$ is dominated by low $\ell'$ values and one expects
$\ell \simeq \ell''$ because of the triangle relation imposed by the 3J symbols
(see Appendix \ref{appendix:3J}).  If one further assumes $C_\ell$ and $b_\ell$
to vary slowly in $\ell$, then Eqs.~(\ref{eq:w3j_ortho1}) and
(\ref{eq:w3j_ortho3}) can be used to impose $s_1+v_1=s_2+v_2=s$ in
Eq.~(\ref{eq:crossCl_general}) and provide
\begin{align}
\pC^{v_1v_2}_{\ell} 
        &= % (-1)^{v_1-v_2} 
        \sum_{u_1u_2} C^{u_1u_2}_{\ell} \frac{k_{u_1}k_{u_2}}{k_{v_1}k_{v_2}}
        \sum_{j_1 j_2} \sum_s
\ _{u_1} \hatb^{(j_1)*}_{\ell,s-v_1} 
\ _{u_2} \hatb^{(j_2) }_{\ell,s-v_2}
\ {\tOmega}^{(j_1j_2)}_{v_1,v_2,s},
\label{eq:cl_arbitrary}
\end{align}
with 
\begin{subequations}
\label{eq:Omega_def}
\begin{align}
        {\tOmega}^{(j_1j_2)}_{v_1,v_2,s} 
        &\equiv 
        \cpem{j_1,v_1}\cpem{j_2,v_2} \frac{1}{4\pi} 
                         \sum_{\ell'm'}      {}_{s}  {\tomega}^{(j_1) }_{\ell'm'}[v_1]
                                      \   {}_{s}  {\tomega}^{(j_2)*}_{\ell'm'}[v_2],
        \label{eq:omega_lm}\\
        &= \cpem{j_1,v_1}\cpem{j_2,v_2} \frac{1}{\npix} 
        \sum_p  
       \tomega_{s}^{(j_1)}[v_1](p)\ 
       \tomega_{s}^{(j_2)*}[v_2](p),
        \label{eq:omega_pixel}\\
        &= {\tOmega}^{(j_1j_2)*}_{-v_1,-v_2,-s}.
        %%%&=& {\tOmega}^{(j_2j_1)*}_{v_2,v_1,s} = {\tOmega}^{(j_2j_1)}_{-v_2,-v_1,-s}
\end{align}
\end{subequations}
%
%==============================
%\onecolumn
%==============================
%
As derived in Appendix \ref{appendix:smooth_scanning}, Eq.~(\ref{eq:cl_arbitrary}) reduces to a mixing equation relating the observed cross-power spectra to the true ones:
\begin{equation}
        \pC^{XY}_{\ell} = \sum_{X'Y'} W^{XY,\; X'Y'}_\ell C^{X'Y'}_{\ell}
        \label{eq:beam_matrix}
\end{equation}
with $X,Y,X',Y' \in \{T,E,B\}$. 

In the smooth scanning case representative of past and forthcoming satellite missions, 
the effect of observing the sky with non-ideal beams is therefore to couple the temperature and polarization 
power spectra $C^{X'Y'}_{\ell}$ at the same multipole $\ell$ through
an extended beam window matrix $W^{XY,\; X'Y'}_\ell$, as illustrated on Fig.~\ref{fig:Wmatrix}.

\subsection{Arbitrary scanning, circular identical beams}
If the scanning beams are now assumed to all be circular and identical, the measured $\pC(\ell)$ will not depend on the details of the scanning strategy, orientation of the detectors, or relative weights of the detectors. We are indeed exactly in the
ideal hypotheses of the map making formalism (Eq.~\ref{eq:ideal_tod}) and get the well known and simple result that the effect of the beam can be factored out.

If one considers detectors with identical circular copolarized beams, and whose actual polarization efficiency was used 
during the map making: $\cpem{j}=\cpei{j}$, such that
\begin{align} 
        {}_u\hatb_{\ell,s}^{(j)} &\equiv w_j q_{\ell}\ _u b_{\ell,s}^{(j)} %e^{i s \gamma_j}
        = w_j q_{\ell} \cpem{j,u} \,  b_{\ell} \, \delta_{s,-u},
\end{align} 
then Eqs.~(\ref{eq:cl_arbitrary}) and (\ref{eq:omega_pixel}) feature terms like
$\sum_{j}\, _{u} \hatb^{(j)*}_{\ell,s-v} \cpem{j,v} \tomega_{s}^{(j)}[v]$, which when written in a matrix form,
verify the equality
\begin{align}
        q_\ell b_\ell \sum_{j} w_j 
        \matrixthree%
        {        {\tomega_{0  }^{(j)}[0]}}
        {\cpem{j}  {\tomega_{ -2}^{(j)}[0] }} %e^{-2i\gamma_j}}}
        {\cpem{j}  {\tomega_{  2}^{(j)}[0] }} %e^{ 2i\gamma_j}}}%%%%
        {\cpem{j}  {\tomega_{  2}^{(j)}[2] }} %e^{ 2i\gamma_j}}}%
        {\cpem{j}^2{\tomega_{0  }^{(j)}[2] }}%
        {\cpem{j}^2{\tomega_{  4}^{(j)}[2] }} %e^{ 4i\gamma_j}}}%%%
        {\cpem{j}  {\tomega_{ -2}^{(j)}[-2]}} %e^{-2i\gamma_j}}}%
        {\cpem{j}^2{\tomega_{ -4}^{(j)}[-2]}} %e^{-4i\gamma_j}}}%
        {\cpem{j}^2{\tomega_{0  }^{(j)}[-2]}}%%%
        & = 
        q_\ell b_\ell \matrixthree%
        {1}{0}{0}{0}{1}{0}{0}{0}{1},
\end{align}
according to Eq.~(\ref{eq:omega0_identity}). The measured power
spectra are then
\begin{equation}
        \pC_{\ell}^{XY} = \hatb_\ell^2 C_\ell^{XY} = q_{\ell}^2 b_\ell^2 C_{\ell}^{XY},
% = \frac{4\pi}{2 \ell+1} b_{\ell}^2 C_{\ell}^{XY}
\end{equation}
and $\pC_{\ell}^{XY} = \exp \left(-\ell(\ell+1) \sigma^2\right)\, C_{\ell}^{XY}$
for the Gaussian circular beam introduced in Eq.~(\ref{eq:gaussbeam_T}).
Obviously, these very simple results assume that the whole sky is observed. 
If not, the cut-sky induced $\ell-\ell$ and $E-B$ coupling effects mentioned at the end of Section \ref{sec:measured_cl} have to be
accounted for, as described, for example, in \citet{Chon+2004}, \citet{Mitra+2009}, \citet{Grain+2009}, and references therein.

%--------------------------
\subsection{Arbitrary beams, ideal scanning}
\label{sec:ideal_scanning}
Let us now consider the case of an ideal scanning of the sky,
for which in any pixel $p$, the number of valid (unflagged) 
samples is the same for all detectors $h_j(p) = h(p)$,
and each detector $j$ covers uniformly all possible orientations within that pixel along the duration of the mission.
This constitutes the ideal limit aimed at by the scanning strategy illustrated in Fig.~\ref{fig:hits_LB}.
The assumption of smooth scanning is then perfectly valid, and details of the calculations can be found in Appendix \ref{appendix:ideal_scanning}. We find for instance that the matrix describing 
how the measured temperature and polarization power spectra are affected by the input $TT$ spectrum reads
\begin{align} 
W_{\ell}^{XY,\; TT} \equiv 
 \left( 
\begin{array}{l}
 W_{\ell}^{TT,\; TT} \vphantom{\hatb_{\ell,0}^{(j_1)*}}\\
 W_{\ell}^{EE,\; TT} \vphantom{\hatb_{\ell,0}^{(j_1)*}}\\
 W_{\ell}^{BB,\; TT} \vphantom{\hatb_{\ell,0}^{(j_1)*}}\\
 W_{\ell}^{TE,\; TT} \vphantom{\hatb_{\ell,0}^{(j_1)*}}\\
 W_{\ell}^{TB,\; TT} \vphantom{\hatb_{\ell,0}^{(j_1)*}}\\
 W_{\ell}^{EB,\; TT} \vphantom{\hatb_{\ell,0}^{(j_1)*}}\\
 W_{\ell}^{ET,\; TT} \vphantom{\hatb_{\ell,0}^{(j_1)*}}\\
 W_{\ell}^{BT,\; TT} \vphantom{\hatb_{\ell,0}^{(j_1)*}}\\
 W_{\ell}^{BE,\; TT} \vphantom{\hatb_{\ell,0}^{(j_1)*}}
\end{array}
 \right)
&= 
\sum_{j_1j_2} 
 \left( 
\begin{array}{l}
%1 ----
 \hatb_{\ell,0}^{(j_2)} \hatb_{\ell,0}^{(j_1)*} 
{\ \myxi_{00}}   \\
%2
 \left( \hatb_{\ell,-2}^{(j_2)}+\hatb_{\ell,2}^{(j_2)} \right)  \left(\hatb_{\ell,-2}^{(j_1)*}+\hatb_{\ell,2}^{(j_1)*}\right) 
{\ \cpem{j_1}\cpem{j_2}\myxi_{22}}  \\
%3
 \left( \hatb_{\ell,-2}^{(j_2)}-\hatb_{\ell,2}^{(j_2)} \right)  \left(\hatb_{\ell,-2}^{(j_1)*}-\hatb_{\ell,2}^{(j_1)*}\right) 
{\ \cpem{j_1}\cpem{j_2}\myxi_{22}} \\
%4 ----
 -\left( \hatb_{\ell,-2}^{(j_2)}+\hatb_{\ell,2}^{(j_2)} \right)  \hatb_{\ell,0}^{(j_1)*} 
{\ \cpem{j_2}\myxi_{02}} \\
%5
 -i \left( \hatb_{\ell,-2}^{(j_2)}-\hatb_{\ell,2}^{(j_2)} \right)  \hatb_{\ell,0}^{(j_1)*} 
{\ \cpem{j_2}\myxi_{02}} \\
%6
 i \left( \hatb_{\ell,-2}^{(j_2)}-\hatb_{\ell,2}^{(j_2)} \right)  \left(\hatb_{\ell,-2}^{(j_1)*}+\hatb_{\ell,2}^{(j_1)*}\right) 
{\ \cpem{j_1}\cpem{j_2}\myxi_{22}} \\
%7 ----
 -\hatb_{\ell,0}^{(j_2)} \left(\hatb_{\ell,-2}^{(j_1)*}+\hatb_{\ell,2}^{(j_1)*}\right) 
{\ \cpem{j_1}\myxi_{20}} \\
%8
 i \hatb_{\ell,0}^{(j_2)} \left(\hatb_{\ell,-2}^{(j_1)*}-\hatb_{\ell,2}^{(j_1)*}\right) 
{\ \cpem{j_1}\myxi_{20}} \\
%9
 -i \left( \hatb_{\ell,-2}^{(j_2)}+\hatb_{\ell,2}^{(j_2)} \right)  \left(\hatb_{\ell,-2}^{(j_1)*}-\hatb_{\ell,2}^{(j_1)*}\right) 
{\ \cpem{j_1}\cpem{j_2}\myxi_{22}} \\
\end{array}
 \right),
 \label{eq:wl_xyTT_is_MainText}
\end{align}
with the normalization factors
\begin{align}
\myxi_{00}^{-1} &= \sum_{k_1k_2}w_{k_1}w_{k_2}                        ,
&              
\myxi_{02}^{-1} &= \sum_{k_1k_2}w_{k_1}w_{k_2}            \cpem{k_2}^2,% \nonumber\\
&
\myxi_{20}^{-1} &= \sum_{k_1k_2}w_{k_1}w_{k_2}\cpem{k_1}^2            ,
&              
\myxi_{22}^{-1} &= \sum_{k_1k_2}w_{k_1}w_{k_2}\cpem{k_1}^2\cpem{k_2}^2.
\end{align}

This confirms that in this ideal case, as expected and discussed previously
 \citep[e.g.,][and references therein]{Wallis+2014}, the leakage from
temperature to polarization (Eq.~\ref{eq:wl_xyTT_is_MainText}) is driven by the beam
ellipticity ($\hatb_{l,\pm2}^{(j)}$ terms) which has the same spin $\pm2$ as
polarization. One also sees that the contamination of the $E$ and $B$ spectra by
$T$ are swapped (e.g., $W_{\ell}^{EE,\; TT} \longleftrightarrow W_{\ell}^{BB,\;
  TT}$) when the beams are rotated with respect to the polarimeter direction by
$45\degr$ ($\hatb_{\ell,\pm2} \longrightarrow \pm i \hatb_{\ell,\pm2}$), as shown in \citet{Shimon+2008}.

%-----------------------------------------------------------------
\begin{figure*}[htbp] 
%-----------------------------------------------------------------
%png 90pt 90  137 137
  \includegraphics[angle=0,width=0.49\textwidth,trim={30pt 250 70 100},clip]{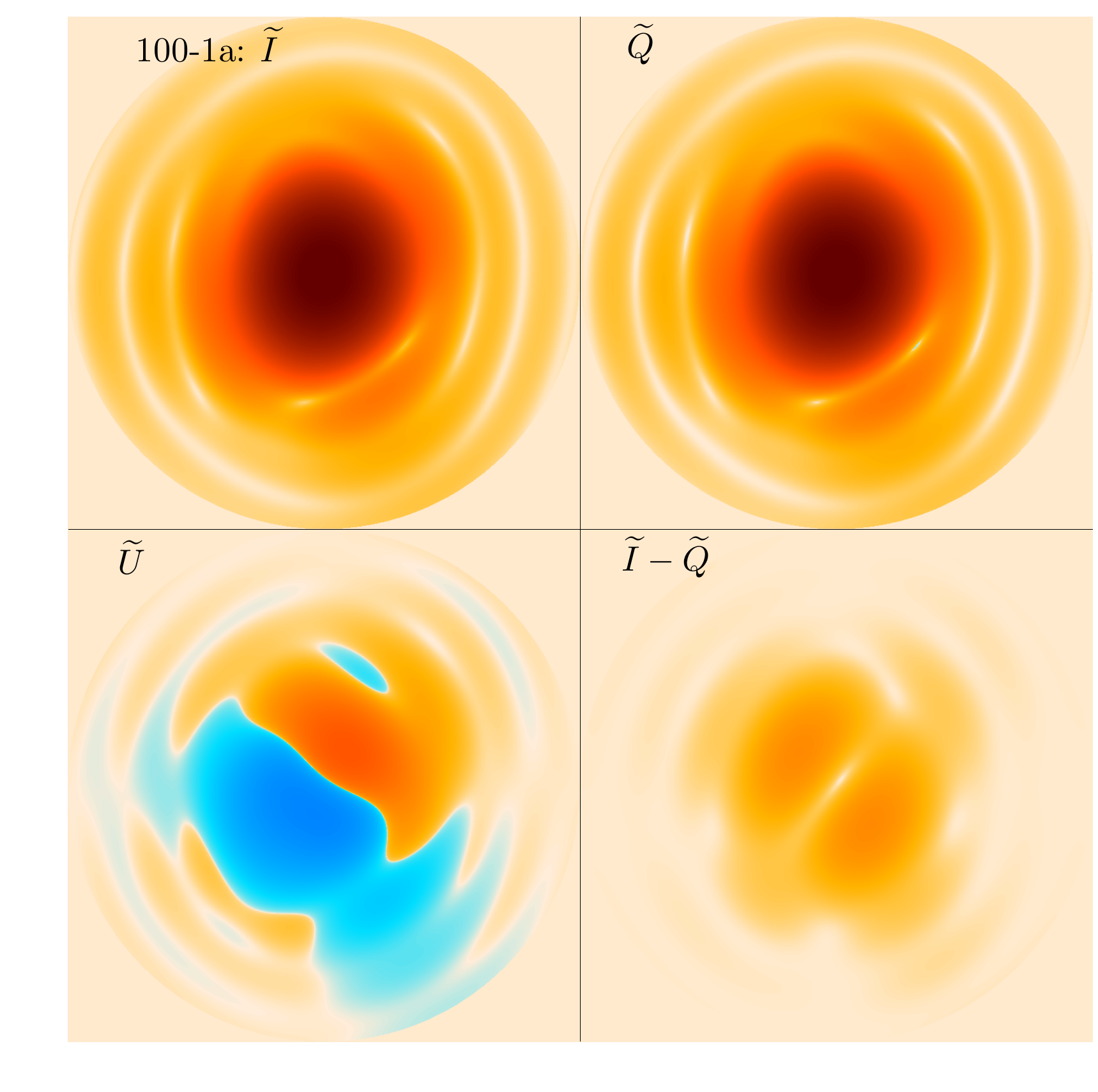}
  \includegraphics[angle=0,width=0.49\textwidth,trim={30pt 250 70 100},clip]{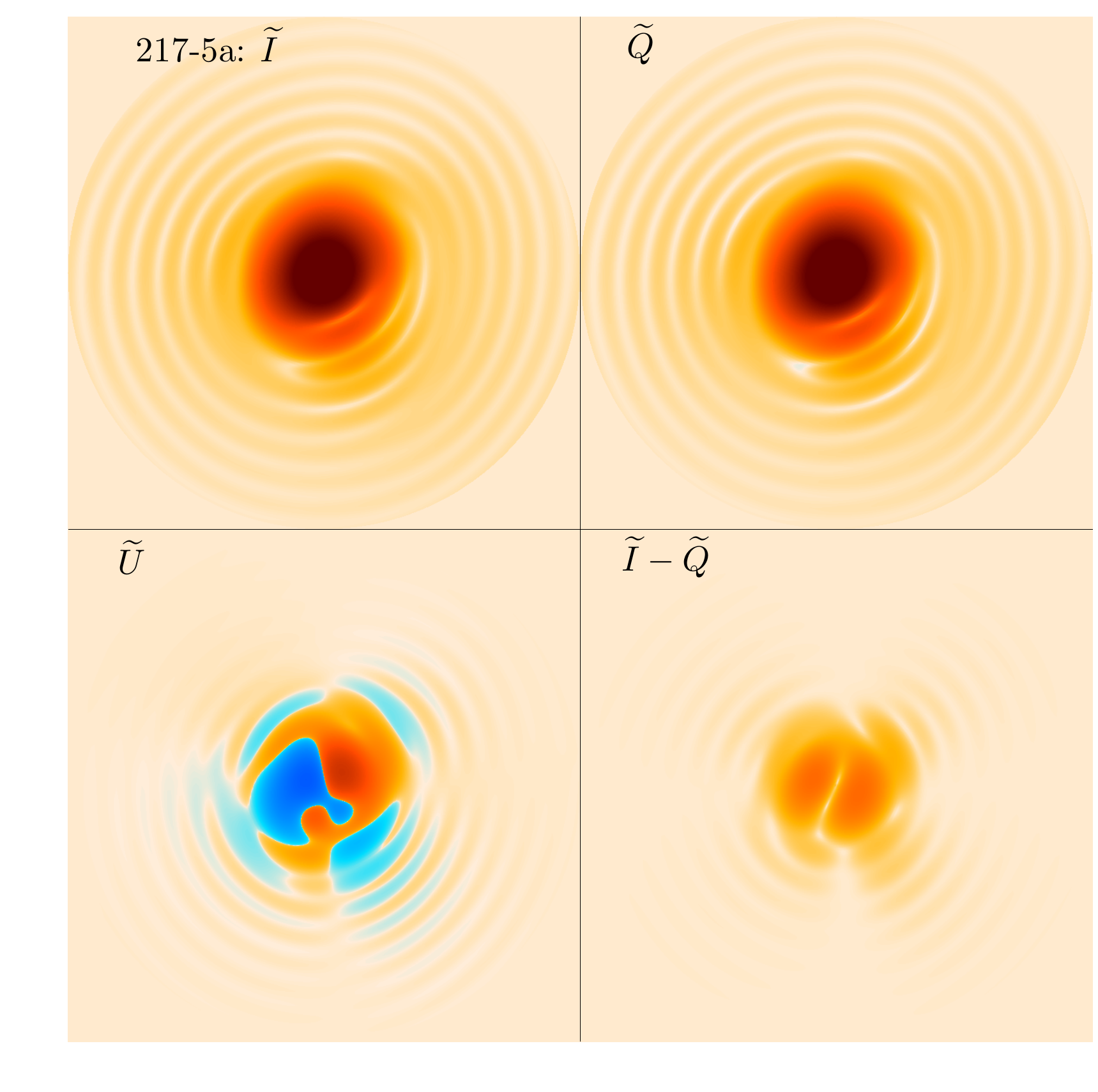}
\begin{center}
  \includegraphics[angle=0,width=0.45\textwidth]{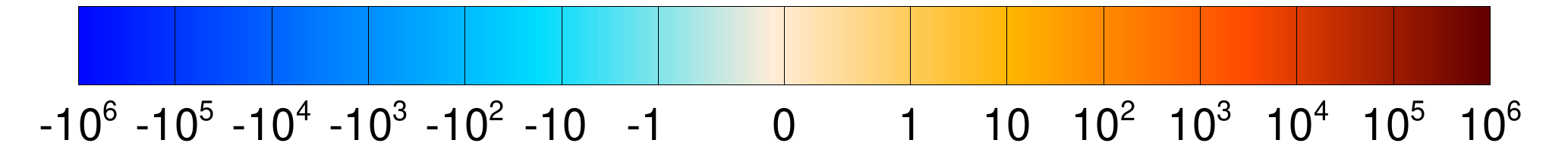}
\end{center}
  \caption{\label{fig:beam_iqu} Computer simulated beam maps 
($\bI$, $\bQ$, $\bI-\bQ$ and $\bU$ clockwise from top-left) for two of the \Planck-HFI detectors (100-1a and 217-5a) used
in the validation of \Quickpol.
Each panel is 1$\degr$x1$\degr$ in size, and the units are arbitrary.}
\end{figure*}
%-----------------------------------------------------------------

\subsection{Finite pixel size and sub-pixel effects}
\label{sec:subpixel}
As shown in \citet{Planck2013-7}, in the case of temperature fluctuations, the
effect of the finite pixel size is twofold. First, in each pixel, the
  distance between the nominal pixel center and the center of mass of the observations 
  couples to the
  local gradient of the Stokes parameters to induce noise terms. Second,
  there is a smearing effect due to the integration of the signal over the
surface of the pixel. Equation~(\ref{eq:beam_matrix}) then becomes
\begin{equation}
        \pC^{XY}_{\ell} = W^{\rm pix}_\ell\, \sum_{X'Y'} W^{XY,\; X'Y'}_\ell C^{X'Y'}_{\ell} + N^{XY}_{\ell}
\end{equation}
with $W^{\rm pix}_\ell = 1 - \ell(\ell+1)\sigma^2/2 + {\cal{O}}\left((\sigma
\ell)^3\right)$, and $\sigma^2 = \VEV{\dd\vecr^2}$ the squared displacement
averaged over the hits in the pixels and over the set of considered pixels.  As
shown in Appendix \ref{sec:append_subpixel}, the additive noise term, sourced by
the temperature gradient within the pixel, affects both temperature and
polarization measurements, with $N^{EE}_{\ell} = N^{BB}_{\ell}$ and $\left|
N^{EE}_{\ell} \right| \la \left| N^{TT}_{\ell} \right|$, while the other spectra
are much less affected, that is, $\left| N^{TE}_{\ell} \right|, \left| N^{TB}_{\ell}
\right|, \left| N^{EB}_{\ell} \right| \ll \left| N^{TT}_{\ell} \right|$. The
sign of this noise term is arbitrary and can be negative when cross-correlating
maps with a different sampling of the pixels.

%==============================
%\twocolumn
%==============================

\section{Numerical implementation}
\label{sec:numerical_implementation}
Numerical implementations of this formalism are performed in three steps,
assuming that the individual beam $b_{\ell s}^{(j)}$ is already computed for
$0\le s \le \smax+4$ and $0 \le \ell \le \lmax$:

\begin{enumerate}
\item For each involved detector $j$, and for $0\le s \le \smax$,
one computes the $s$-th complex moment of its direction of polarization in
pixel $p$: $\omega_s^{(j)}(p)$ defined in Eq.~(\ref{eq:spin_omega}).  
Since this
requires processing the whole scanning data stream, this step can be time
consuming. However it has to be computed only once for all cases, independently of
the choices made elsewhere on the beam models, calibrations, noise weighting, and other factors. As we
shall see below, it may not even be necessary to compute it, or store
it, for every sky pixel.

\item The $\omega_s^{(j)}(p)$ computed above are weighted with the assumed
  inverse noise variance weights $w_j$ and polar efficiencies $\cpem{j}$ to
  build the hit matrix $\matH$ in each pixel, which is then inverted to compute
  the $\tomega_s^{(j)}(p)$, defined in Eq.~(\ref{eq:define_tildeomega}).  Those
  are then multiplied together to build the scanning information matrix
  $\widetilde{\bf \Omega}$ using its pixel space definition
  (Eq.~\ref{eq:omega_pixel}). The resulting complex matrix contains
  $9n_{1}n_{2}(2\smax+1)$ elements, where $n_1$ and $n_2$ are the number of
  detectors in each of the two detector assemblies whose cross-spectra are
  considered.  This step can be parallelized to a large extent, and can be
  dramatically sped up by building this matrix out of a representative subset of
  pixels. In our comparison to simulations, described in Section
  \ref{sec:simulations}, and performed on HEALPix map with $\nside=2048$ and
  $\npix = 12 \nside^2 = 50\, 10^6$ pixels, we checked that using only
  $\npix/64$ pixels evenly spread on the sky gave final results almost identical
  to those of the full calculations.

\item Finally, using Eqs.~(\ref{eq:main}-\ref{eq:beam_matrix}) we
  note that $W^{XY,\; X'Y'}_\ell = \partial \pC^{XY}_{\ell} / \partial
  C^{X'Y'}_{\ell}$, so that, for instance, for a given $\ell$, the 3x3 $W^{XY,\;
    TE}_\ell$ matrix is computed by replacing in Eq.~(\ref{eq:main}) its central
  term $\matC_{\ell}$ with its partial derivative, such as
\begin{align}
\frac{\partial}{\partial C_\ell^{TE}}
\matC_{\ell}
& =
\matR_2.
\frac{\partial}{\partial C_\ell^{TE}}
\matrixthree%
{C_{\ell}^{TT}}%
{C_{\ell}^{TE}}%
{C_{\ell}^{TB}}%
{C_{\ell}^{ET}}%
{C_{\ell}^{EE}}%
{C_{\ell}^{EB}}%
{C_{\ell}^{BT}}%
{C_{\ell}^{BE}}%
{C_{\ell}^{BB}}%
.\matR_2^{\dagger} \\
& = 
\matR_2.
\matrixthree%
{0}%
{1}%
{0}%
{1}%
{0}%
{0}%
{0}%
{0}%
{0}%
.\matR_2^{\dagger},
\label{eq:Clmatrix_derivative}
\end{align}
where we assumed in Eq.~(\ref{eq:Clmatrix_derivative}) that, on the
sky, ${C_{\ell}^{TE}}={C_{\ell}^{ET}}$ and generally
${C_{\ell}^{X'Y'}}={C_{\ell}^{Y'X'}}$, like for CMB anisotropies.

On the other hand, when dealing with arbitrary foregrounds cross-frequency
spectra, we would have to assume ${C_{\ell}^{X'Y'}} \neq {C_{\ell}^{Y'X'}}$ when
$X' \neq Y'$, and compute $W^{XY,\; X'Y'}_\ell$ and $W^{XY,\; Y'X'}_\ell$
separately. As we shall see in Section \ref{sec:propagation}, this final and
fastest step is the only one that needs to be repeated in a Monte-Carlo
analysis of instrumental errors, and it can be sped up. Indeed,
since the input $b_{\ell m}$ and output $W_\ell$ are generally very smooth
functions of $\ell$, it is not necessary to do this calculation for every single
$\ell$, but rather for a sparse subset of them, for instance regularly
interspaced by $\delta\ell$. The resulting $W_\ell$ matrix is then B-spline
interpolated. In our test cases with $\fwhm = 10$ to $5'$, using $\delta\ell=10$
leads to relative errors on the final product below $10^{-5}$ for each $\ell$.

\end{enumerate}
In our tests, with $\smax=6$, $\lmax=4000$, $n_1=n_2=4,$ and all proposed
speed-ups in place, Step 2 took about ten minutes, dominated by IO, while Step 3 took
less than a minute on one core of a 3GHz Intel Xeon CPU. The final product is a set of six
(or nine) real matrices $W^{XY,\; X'Y'}_\ell$, each with $9(\lmax+1)$ elements.

%-----------------------------------------------------------------
\begin{figure*}[htbp]
%-----------------------------------------------------------------
  \includegraphics[angle=0,width=0.49\textwidth]{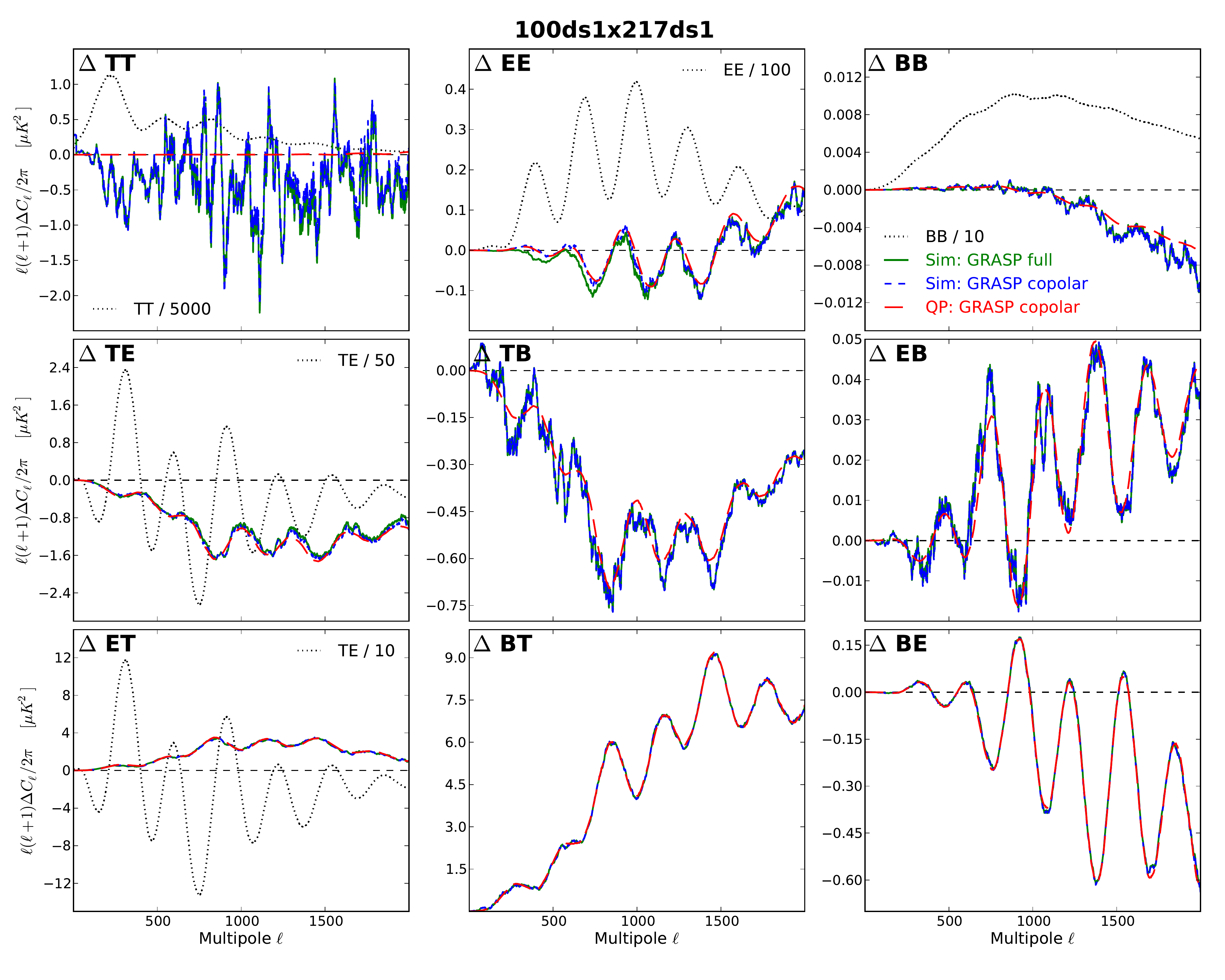}
  \includegraphics[angle=0,width=0.49\textwidth]{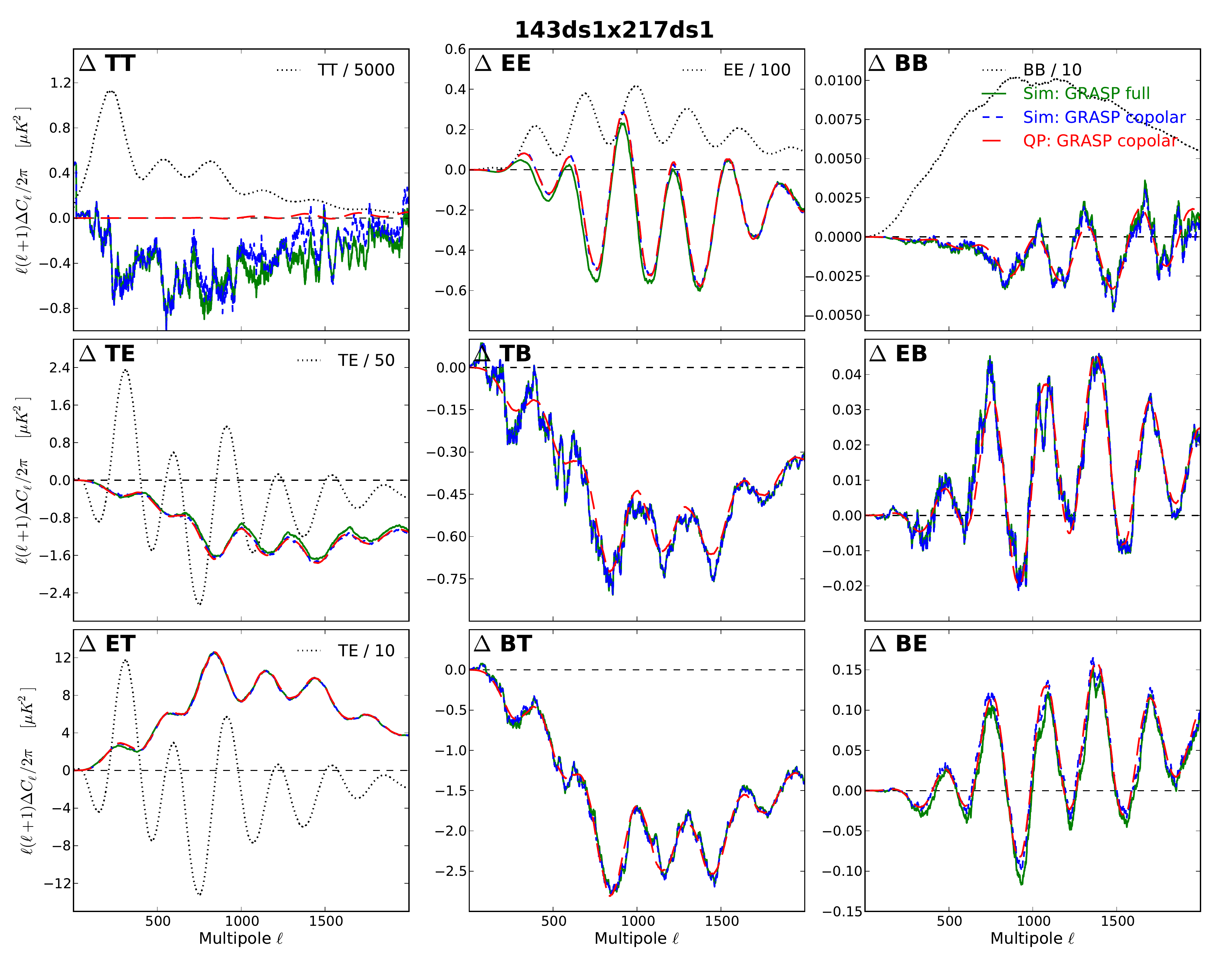}
  \caption{\label{fig:sim_deltacl} Comparison to simulations for 
100ds1x217ds1 (lhs panels)   %\onetwocols{\cols}{({\em lhs} panels)}{(upper panels)}
and 
143ds1x217ds1 (rhs panels)   %\onetwocols{\cols}{({\em rhs} panels)}{(lower panel)}
 cross power spectra, for computer simulated beams. In each panel is shown the discrepancy between the actual $\ell(\ell+1)C_{\ell} / 2\pi$ and the one in input, smoothed on $\Delta \ell = 31$. Results obtained on simulations with either the full beam model (green curves) or the co-polarized beam model (blue dashes) are to be compared to \Quickpol{} analytical results (red long dashes). In panels where it does not vanish, a small fraction of the input power spectrum is also shown as black dots for comparison.}
\end{figure*}
%-----------------------------------------------------------------
%======================================================================
\section{Comparison to \Planck-HFI simulations}
\label{sec:simulations}
The differential nature of the polarization measurements, in the absence of
modulating devices such as rotating half-wave plates, means that any mismatch
between the responses of the two (or more) detectors being used will leak a
fraction of temperature into polarization. This was observed in \Planck, even
though pairs of polarized orthogonal detectors observed the sky through the same
horn, therefore with almost identical optical beams. Optical mismatches within pairs of
detectors were enhanced by residuals of the electronic time response
deconvolution which could affect their respective scanning beams differently
\citep{Planck2013-4, Planck2013-7}. Other sources of mismatch included their 
different noise levels and thus their respective
statistical weight on the maps, which could reach relative differences of up to 80\%, 
and the number of valid samples which could vary by up to 20\% between detectors.
As seen previously, these detector-specific features can be included in the \Quickpol{} pipeline in order
to describe as closely as possible the actual instrument. In this section,
  we show how we actually did it and how \Quickpol{} compares to full-fledged
  simulations of \Planck-HFI observations.

Noiseless simulations of \Planck-HFI observations of a pure CMB sky were run for
quadruplets of polarized detectors at three different frequencies (100, 143, and
217GHz), and identified as 100ds1, 143ds1, and 217ds1 respectively.  The input
CMB power spectrum $C^{XY}_\ell$ was assumed to contain no primordial tensorial
modes, with the traditional $C^{TB}_\ell = C^{EB}_\ell = 0$ and $C^{XY}_\ell =
C^{YX}_\ell$.  The same mission duration, pointing, polarization orientations
($\gamma_j$) and efficiencies ($\cpem{j}$), flagged samples, and discarded
pointing periods ($f_j$) were used as in the actual observations, with computer
simulated polarized optical beams for the relevant detectors produced with the
GRASP\footnote{TICRA: \url{http://www.ticra.com}.} physical optics code
\citep[][and references therein]{Rosset+2007} as illustrated on
Fig.~\ref{fig:beam_iqu}.  Data streams were generated with the \mycode{LevelS}
simulation pipeline \citep{LevelS2006}, using the \mycode{Conviqt} code
\citep{Conviqt2010} to perform the convolution of the sky with the beams,
including the $b_{\ell s}$ for $|s|\le \smax = 14$ and $\ell \le \lmax=4800$.
%http://pipelines.planck.fr/pipeline_id.php?db=simdb&id=1010122858
Polarized maps of each detector set were produced with the \mycode{Polkapix}
destriping code \citep{Polkapix2011}, assuming the same noise-based relative
weights ($w_j$) as the actual data, and their cross spectra were computed over
the whole sky with HEALPix \mycode{anafast} routine to produce the empirical
power spectra $\hC^{XY}_\ell$.\\ The same exercise was reproduced replacing the
initial $\bI,\bQ,\bU$ beam maps with a purely co-polarized beam based on the same
$\bI$, in order to test the validity of the co-polarized  assumption in \Planck.

Figure \ref{fig:sim_deltacl} shows how the empirical power spectra are different from the input ones,
\begin{align}
        \Delta \hC^{XY}_\ell &= 
        \frac{\hC^{XY}_\ell}{%
        W^{\rm pix}_\ell W^{XY,\;XY}_\ell} - C^{XY}_\ell,
\end{align}
after correction from the pixel and (scalar) beam window functions, 
and compares those to the \Quickpol{} predictions 
\begin{align}
        \Delta \pC^{XY}_\ell &= 
        \frac{\displaystyle W^{\rm pix}_\ell \sum\limits_{X'Y'} W^{XY,\; X'Y'}_\ell C^{X'Y'}_\ell}{%
        W^{\rm pix}_\ell W^{XY,\;XY}_\ell} - C^{XY}_\ell,
\end{align}
for all nine possible values of $XY$ for the cross-spectra of detector sets
100ds1x217ds1 and 143ds1x217ds1.  The results are actually multiplied by the
usual $\ell(\ell+1)/ 2\pi$ factor, and smoothed on $\Delta \ell = 31$.  The
empirical results are shown both for the full-fledged beam model (green curves)
and the purely co-polarized model (blue dashes).  One sees that the change, mostly
visible in the $EE$ case, is very small, validating the co-polarized beam
assumption, at least within the limits of this computer simulated \Planck{}
optics.  The \Quickpol{} predictions, only shown in the co-polarized case for clarity
(long red dashes), agree extremely well with the corresponding numerical
simulations. We have checked that this agreement to simulations remains true in
the full beam model.

%======================================================================
\section{Propagation of instrumental uncertainties}
\label{sec:propagation}
We assumed so far the instrument to be non-ideal, but exactly known. In practice, however,
the instrument is only known with limited accuracy and the final beam matrix
will be affected by at least four types of uncertainties:
\begin{itemize}
\item limited knowledge of the beam angular response, which affects the
  $b^{(j)}_{\ell m}$, replacing them with $b'^{(j)}_{\ell m}$ while preserving
  the beam total throughput after calibration (see below)
  $b'^{(j)}_{00}=b^{(j)}_{00}$.  We therefore assume the beam power spectrum
  $W_\ell = \sum_m |b_{\ell m}|^2 / (2\ell+1)$ to be the same at $\ell=0$, where
  the beam throughput is defined, and at $\ell=1$, where the detector gain
  calibration is usually done using the CMB dipole.
\item error on the gain calibration of detector $j$, which translates into 
  $b_{\ell m}^{(j)} \longrightarrow (1 + \delta c_j) \ b_{\ell m}^{(j)}$, with $|\delta c_j| \ll 1$,
\item error on the polar efficiency of detector $j$, which translates into 
  $\cpei{j} = \cpem{j} (1+\delta \cpem{j}/\cpem{j})$. 
  As discussed in Section~\ref{sec:datastream}, 
  we expect in the case of \Planck-HFI a relative uncertainty $|\delta \cpem{j}/\cpem{j}| < 1\%$.
\item error on the actual direction of polarization: for each detector $j$, the
  direction of polarization measured in a common referential becomes $\gamma_j
  \longrightarrow \gamma_j + \delta \gamma_j$. In the case of \Planck-HFI,
  \citet{Rosset+2010} found the pre-flight measurement of this angle to be
  dominated by systematic errors of the order of $1\degr$ for polarization sensitive bolometers (PSBs). These
  uncertainties can be larger for spider web bolometers (SWBs), 
 but as we shall see below, the coupling
  with the low polarization efficiency $\cpem{j}$ of those detectors makes them
  somewhat irrelevant.
\end{itemize}
All these uncertainties can be inserted in Eq.~(\ref{eq:main}) by substituting Eq.~(\ref{eq:beam_mat1})
\begin{equation}
\hat{\matB}^{(j)}_{\ell,s} =
\matrixthree%
{        \hatb^{(j)}_{\ell,s  }}%
{        \hatb^{(j)}_{\ell,s-2}}%
{        \hatb^{(j)}_{\ell,s+2}}%
{\cpem{j}\hatb^{(j)}_{\ell,s+2}}%
{\cpem{j}\hatb^{(j)}_{\ell,s  }}%
{\cpem{j}\hatb^{(j)}_{\ell,s+4}}%
{\cpem{j}\hatb^{(j)}_{\ell,s-2}}%
{\cpem{j}\hatb^{(j)}_{\ell,s-4}}%
{\cpem{j}\hatb^{(j)}_{\ell,s  }},%
\nonumber
\end{equation}
with
\begin{equation}
\hat{\matB}'^{(j)}_{\ell,s} =
(1 + \delta c_j)
\matrixthree%
{1}{0}{0}%
{0}{\cpem{j}x_j}{0}%
{0}{0}{\cpem{j}x_j^*}
.
\matrixthree%
{\hatb'^{(j)}_{\ell,s  }}%
{\hatb'^{(j)}_{\ell,s-2}}%
{\hatb'^{(j)}_{\ell,s+2}}%
{\hatb'^{(j)}_{\ell,s+2}}%
{\hatb'^{(j)}_{\ell,s  }}%
{\hatb'^{(j)}_{\ell,s+4}}%
{\hatb'^{(j)}_{\ell,s-2}}%
{\hatb'^{(j)}_{\ell,s-4}}%
{\hatb'^{(j)}_{\ell,s  }},%
\label{eq:beam_matrix_error}
\end{equation}
where $x_j=(1+\delta\cpem{j}/\cpem{j}) e^{2i\delta\gamma_j}$.

As mentioned in Section~\ref{sec:results}, such substitutions are done in Step 3 of the \Quickpol{} pipeline. A new set of numerical values for the instrument model can therefore be turned rapidly into a beam window matrix (Eq.~\ref{eq:beam_matrix}), allowing, for instance, a Monte-Carlo exploration at the power spectrum level of the instrumental uncertainties.

%======================================================================
\section{About rotating half-wave plates}
\label{sec:rhwp}
In the previous sections we have focused on experiments that rely on the
rotation of the full instrument with respect to the sky to have the angular redundancy
required to measure the Stokes parameters. An alternative way is to rotate the
incoming polarization at the entrance of the instrument while leaving the rest
fixed. This is most conveniently achieved with a rotating half-wave plate
(\rhwp). The rotation is either stepped \citep{polarbear_2014} or continuous
\citep{ebex_2014,abs_2016,Ritacco+2016a}. The advantages of this system are
numerous, the first of which is the decoupling between the optimization of the
scanning strategy in terms of ``pure'' redundancy and its optimization in
terms of ``angular'' redundancy. It is much easier to control the rotation of
a \rhwp{} than of a full instrument and therefore ensure an optimal angular
coverage whatever the observation scene is. If the rotation is continuous and
fast, typically of the order of 1~Hz, it has the extra advantage of modulating
polarization at frequencies larger than the atmospheric and electronics $1/f$ noise
knee frequency, hence ensuring a natural rejection of these low frequency
noises. Furthermore, this allows us to build $I$, $Q$ , and $U$ maps per detector,
without needing to combine different detectors with their associated bandpass
mismatch or other differential systematic effects mentioned in the previous
sections. Individual detector systematics therefore tend to average out rather than
combine to induce leakage between sky components. On the down side, this comes at the
price of moving a piece of hardware in the instrument and all its associated
systematic effects, starting with a signal that is synchronous with the \rhwp{}
rotation as observed in \citet{Johnson+2007}, \citet{ebex_2014}, and \citet{Ritacco+2016b}. \\
Such trade-offs are being
investigated by current experiments using {\rhwp}s and will certainly be studied
in more details in preparation of future CMB orbital and sub-orbital missions, 
such as CMB-S4 network \citep{cmbs4_2016}. We here
briefly comment on how the addition of a \rhwp{} to an instrument can be coped with
in \Quickpol{}.

The Jones matrix of a HWP (which shifts the $y$-axis electric field by a half period) 
rotated by an angle $\psi$ is \citep{ODea+2007}
\begin{align}
        \matJ_{\rm \rhwp}(\psi) & = \matR_\psi. \jm{1}{0}{0}{-1}.\matR^\dagger_\psi, \\
                         & = \jm{\cos 2\psi}{\sin 2\psi}{\sin 2\psi}{-\cos 2\psi}.
\end{align}
If a rotating HWP is installed at the entrance of the optical system, the Jones
matrix of the system becomes $\matJ\left(\vecr_\alpha\right) \longrightarrow
\matJ\left(\vecr_\alpha,\psi\right) = \matJ\left(\vecr_\alpha\right) \matJ_{\rm
  \rhwp}(\psi)$, and the signal observed in the presence of arbitrary beams
(Eq.~\ref{eq:datastream_beam}) becomes (after dropping the circular polarization
$V$ terms)
\begin{align}
        d(\alpha,\psi) &= \frac{1}{2} \int {\rm d}\vecr 
        \left[ 
         \bI (\alpha, \psi, \vecr)T(\vecr)
        +\bQ (\alpha, \psi, \vecr)Q(\vecr)
        +\bU (\alpha, \psi, \vecr)U(\vecr)
        %-\bV (\alpha, \vecr)V(\vecr) 
        \right],
        \label{eq:datastream_beam_rHWP}
\end{align}
with
\begin{subequations}
\label{eq:beamstokesrotation_rHWP}
\begin{align}
\bI (\alpha, \psi, \vecr) &= \bI(\vecr_\alpha), \\
\bQ (\alpha, \psi, \vecr) &= \bQ(\vecr_\alpha)\cos (2\alpha+4\psi) + \bU(\vecr_\alpha)\sin (2\alpha+4\psi), \\
\bU (\alpha, \psi, \vecr) &= \bQ(\vecr_\alpha)\sin (2\alpha+4\psi) - \bU(\vecr_\alpha)\cos (2\alpha+4\psi).
%, \\
%\bV (\alpha, \vecr) &= \bV(\vecr_\alpha).
\end{align}
\end{subequations}

These new beams can then be passed to Eq.~(\ref{eq:ideal_tod_matrix}) and
  propagated through the rest of \Quickpol{}. Together with
Eq.~(\ref{eq:beamstokesrotation}), we see that, if $\psi$ is correctly chosen, the
modulation of $Q$ and $U$, by $2\alpha+4\psi$, is now clearly different from
that of $T$ which depends only on $\alpha$ via $\vecr_\alpha$, even for
non-circular $\bI$ beams. The leakages from temperature to polarization are
therefore expected to be much smaller than when the polarization modulation is
performed only by a rotation of the whole instrument, and \citet{ODea+2007}
showed, that even for non-ideal \rhwp, the induced systematic effects are
limited to polarization cross-talks without temperature to polarization leakage.

 As previously mentioned, specific systematic effects such as the rotation
  synchronous signal must be treated with care. 
Once such time domain systematic effects are identified and modeled, 
they, together with realistic optical properties of the instrument, can be integrated in the \Quickpol{}
formalism in order
to be taken into account, quantified, and/or marginalized over at the power spectrum level.

\section{Conclusions}
\label{sec:conclusions}

 Polarization measurements are mostly obtained by differencing observations
  by different detectors. Mismatch in their optical beams, time responses,
  bandpasses, and so on induces systematic effects, for example,~temperature to
  polarization leakage. The \Quickpol{} formalism allows us to compute
accurately and efficiently the induced cross-talk between
temperature and polarization power spectra. It also provides a fast and easy
way to propagate instrumental modeling uncertainties down to the final
angular power spectra and is thus a powerful tool to simulate observations 
and to help with the design and specifications of 
  future experiments, such as acceptable beam
  distortions, polarization modulation optimization, and observation
  redundancy. It can cope with time varying instrumental parameters, realistic
  sample flagging, and rejection. The method was validated through comparison to
numerical simulations of realistic \Planck{} observations.  The hypotheses
required on the instrument and survey, described in Sections~\ref{sec:formalism}
and \ref{sec:results}, are extremely general and apply to \Planck{} and to
forthcoming CMB experiments such as PIXIE, LiteBIRD, COrE, and others.  Contrary to
Monte-Carlo based methods, such as \mycode{FEBeCoP}, the impact of the beam
related imperfections on the measured power spectra are obtained without having
to assume any prior knowledge of the sky power spectra.

Of course, the beam matrices provided by \Quickpol{} can be used in the
cosmological analysis of a CMB survey. Indeed, the sky power spectra can
  be modeled as functions of cosmological parameters $\{\theta_C\}$, foreground
  modeling $\{\theta_F\}$ , and nuisance parameters $\{\theta_n\}$. These
  $C^{XY}_\ell(\{\theta_C\}, \{\theta_F\}, \{\theta_n\})$ can then be generated,
  multiplied with the beam matrices $W^{X'Y',\;XY}_\ell$ for the set of
  detectors being analyzed, and compared to the measured $\pC^{X'Y'}_\ell$ in a
  maximum likelihood sense, in the presence of instrumental noise.
The parameters $\{\theta_C\}, \{\theta_F\}, \{\theta_n\}$ can be iterated or integrated upon,
with statistical priors, until a posterior distribution is built. In this kind of forward
approach, it is not necessary to correct the observations from possibly
singular transfer functions, nor to back-propagate the noise. At least some of
the instrumental uncertainties $\{\theta_I\}$ affecting the effective beam via
$W^{X'Y',\;XY}_\ell(\{\theta_I\})$ could be included in the overall analysis, and
marginalized over, thanks to the fast calculation times by \Quickpol{} of the impact of
changes in the gain calibrations, polarization angles, and
efficiencies, as discussed in Section~\ref{sec:propagation}. While \Quickpol{} has been originally developed and tested
  in the case of experiments without a rotating half-wave plate, it is
  straightforward to add one to the current pipeline and assess its impact on
  the aforementioned systematics. Specific additional effects such as a HWP
  rotation synchronous signal or the effect of a tilted HWP are expected to show
  up in real experiments. As long as these can be physically modeled, they can
  be inserted in \Quickpol{} as well.

%------------------------------------------------------------------------------------------
%------------------------------------------------------------------------------------------
\begin{acknowledgements}
Thank you to the \Planck{} collaboration, and in particular to 
D. Hanson, K. Benabed, and F. R. Bouchet for fruitful discussions.
Some results presented here were obtained with the {\tt HEALPix} library.% and {\tt Mathematica}.
\end{acknowledgements}
%------------------------------------------------------------------------------------------
%------------------------------------------------------------------------------------------
%\bibliographystyle{myaat} % long format (title + link to arxiv)
\bibliographystyle{aat} % long format (title + link to arxiv)
\bibliography{quickpol_v2_bib}% no space between bib files
%------------------------------------------------------------------------------------------
%------------------------------------------------------------------------------------------

%\newpage
%==========================================================================
%==========================================================================
%==========================================================================
%==========================================================================
%==========================================================================
%==========================================================================
%==========================================================================
%==========================================================================
%==========================================================================
%==========================================================================
%==========================================================================
%==========================================================================
\appendix

%---------------------------------------------
\section{Projection of maps on spherical harmonics}
\label{appendix:projection_sh}

Here we give more details on the steps required to go from Eq.~(\ref{eq:tvmp}) to
Eq.~(\ref{eq:crossCl_general}). Let us recall Eq.~(\ref{eq:tvmp}) and explain it further:

\begin{align}%---------
\tvm(p) &\equiv \vectorthree{\tm(0;p)}{\tm(2;p)/2}{\tm(-2;p)/2}, \\
         & = \left(\sum_k \sum_{t\in p} A^{(k)\dagger}_{p,t} w_k f_{k,t} A^{(k)}_{t,p} \right)^{-1} 
                 \left(\sum_j \sum_{t\in p} A^{(j)\dagger}_{p,t} w_j f_{j,t}
                 d_{j,t} \right),\nonumber\\
        &= \left(  
                \sum_k w_k
 \matrixthree{%
          \omega_{0}^{(k)}}
 {\cpem{k}  {\omega_{-2}^{(k)}}} %e^{-2i\gamma_k}}}
 {\cpem{k}  {\omega_{ 2}^{(k)}}} %e^{ 2i\gamma_k}}}%%%%
 {\cpem{k}  {\omega_{ 2}^{(k)}}} %e^{ 2i\gamma_k}}}%
 {\cpem{k}^2{\omega_{ 0}^{(k)}}}%
 {\cpem{k}^2{\omega_{ 4}^{(k)}}} %e^{ 4i\gamma_k}}}%%%
 {\cpem{k}  {\omega_{-2}^{(k)}}} %e^{-2i\gamma_k}}}%
 {\cpem{k}^2{\omega_{-4}^{(k)}}} %e^{-4i\gamma_k}}}%
 {\cpem{k}^2{\omega_{ 0}^{(k)}}}%%%
        \right)^{-1} 
 \onetwocols{\cols}{}{\nonumber \\  &  \mydot}
 \sum_{j\ell ms} w_j (-1)^s q_{\ell}\ _{-s}Y_{\ell m}(p)
 \vectorthree%
 {      \omega_{s}^{(j)}}%
 {\cpem{j}\omega_{s+2}^{(j)}} %e^{ 2i\gamma_j}}%
 {\cpem{j}\omega_{s-2}^{(j)}} %e^{-2i\gamma_j}}
 \vectorthree%
 {_0 b^{(j)*}_{\ell s}}%
 {_2 b^{(j)*}_{\ell s}}%
 {_{-2} b^{(j)*}_{\ell s}}^T
 \vectorthree%
 {_0 a_{\ell m}}%
 {_2 a_{\ell m}/2}%
 {_{-2} a_{\ell m}/2}, \nonumber \\
        &=
 \sum_{j\ell ms} (-1)^s \ _{-s}Y_{\ell m}(p)
 \vectorthree%
 {      \tomega_{s}^{(j)}}%
 {\cpem{j}\tomega_{s+2}^{(j)}} %e^{ 2i\gamma_j}}%
 {\cpem{j}\tomega_{s-2}^{(j)}} %e^{-2i\gamma_j}}
 \vectorthree%
 {_0    \hatb^{(j)*}_{\ell s}}%
 {_2    \hatb^{(j)*}_{\ell s}}%
 {_{-2} \hatb^{(j)*}_{\ell s}}^T
 \vectorthree%
 {_0 a_{\ell m}}%
 {_2 a_{\ell m}/2}%
 {_{-2} a_{\ell m}/2},%\label{eq:map_making_matrix}
 \end{align}%---------
 where we introduced the $s$-th complex moment of the direction of
 polarization for detector $j$,
\begin{equation}
        \omega_s^{(j)}(p) \equiv \sum_{t\in p} f_{j,t} e^{i s \alpha^{(j)}_t},
        \label{eq:spin_omega}
\end{equation}
the hit matrix $\matH$ defined for $(u,v) \in \{0,2,-2\}^2$ as
\begin{equation}
        H_{vu}(p) \equiv \sum_j w_j\ \omega_{v-u}^{(j)}(p) %e^{i(v-u)\gamma_j} 
        \cpem{j,v} \cpem{j,u},
        \label{eq:cov_mat}
\end{equation}
with
\begin{equation}
  \cpem{j,v} \equiv \delta_{v,0} + \cpem{j} \left(\delta_{v,-2}+\delta_{v,2}\right),
\end{equation}
the hit normalized moments
\begin{equation}
\vectorthree%
{        \tomega_{s  }^{(j)}(p)}%
{\cpem{j}\tomega_{s+2}^{(j)}(p)} %e^{ 2i\gamma_j}}%
{\cpem{j}\tomega_{s-2}^{(j)}(p)} %e^{-2i\gamma_j}}
\equiv \matH(p)^{-1} 
\vectorthree%
{        \omega_{s  }^{(j)}(p)}%
{\cpem{j}\omega_{s+2}^{(j)}(p)} %e^{ 2i\gamma_j}}%
{\cpem{j}\omega_{s-2}^{(j)}(p)} %e^{-2i\gamma_j}},
\label{eq:define_tildeomega}
\end{equation}
which are described in Appendix \ref{appendix:hit_matrix}, and finally the inverse noise variance weighted beam spherical harmonics (SH) coefficients
\begin{align}
        \ _{u}\hatb_{\ell,s}^{(j)} &\equiv w_j q_{\ell}\ _{u}b_{\ell,s}^{(j)} %e^{i s \gamma_j}
        \label{eq:def_hatb}
    ,\\
                                     &=  \cpei{j}\hatb_{\ell,s+u}^{(j)}.
\end{align}
Since the solution of Eq.~(\ref{eq:map_making_top}) remains the same
when all the noise covariances are rescaled simultaneously by an arbitrary
factor $a$: $\matN_j \longrightarrow a\,\matN_j$, one can also rescale the
weights $w_j$ appearing in Eqs~(\ref{eq:cov_mat}) and (\ref{eq:def_hatb}), with for instance $w_j \longrightarrow w_j/\sum_k w_k$ without altering the final result.

The components of the observed polarized map are then
\begin{align}
\tm(v;p) =& \sum_u \frac{k_u}{k_v} \sum_j  \sum_s \cpem{j,v}\,\tomega_{s+v}^{(j)}(p) %e^{i v \gamma_j}
\onetwocols{\cols}{}{\nonumber \\ & \times}
        \sum_{\ell m}  \ _u a_{\ell m} \ _u \hatb^{(j)*}_{\ell s} (-1)^s \ _{-s}Y_{\ell m}(p),
        \label{eq:mapcomponents}
\end{align}
with
\begin{equation}
        k_0 = 1,\quad k_{\pm 2}=1/2.
\end{equation}
%
% Using $_{s}Y^*_{\ell m} = (-1)^{s+m}\ _{-s}Y_{l-m}$, we note that $\mu(v,u) =
% \mu(-v,-u)^*$.
%
After expansion of the hit normalized moments (Eq.~\ref{eq:define_tildeomega}) 
in spherical harmonics:
\begin{equation}
        \tomega_{s+v}^{(j)}(p) = 
\sum_{\ell'm'} \ _{s+v} \tomega^{(j)}_{\ell'm'} \ _{s+v} Y_{\ell'm'}(p),
\end{equation}
the polarized map reads
\begin{align}
\tm(v;p) &=  \sum_u \frac{k_{u}}{k_{v}} \sum_{j \ell m s \ell' m'} (-1)^s 
        \ _{-s}Y_{\ell m}(p) \ _{s+v} Y_{\ell'm'}(p) 
\onetwocols{\cols}{}{\nonumber \\ & \quad \quad \times}
        \ _u a_{\ell m} \ _u \hatb^{(j)*}_{\ell s} \  %e^{i v \gamma_{j}} 
        \cpem{j,v}\ _{s+v}\tomega^{(j)}_{\ell'm'},
\end{align} %%%%%%%
and the SH coefficients of spin $x$ of map $\tm(v;p)$  are, for pixels of area $\Omega_p$,
\begin{align}
\ _x \tm_{\ell''m''}(v) &\equiv \sum_p \Omega_p  \tm(v;p)\  _{x}Y^*_{\ell''m''}(p)
         = \int \dd\vecr\  \tm(v;\vecr)\  _{x}Y^*_{\ell''m''}(\vecr) \\
   &= \sum_u \frac{k_{u}}{k_{v}} \sum_j %e^{i v \gamma_{j}} 
        \sum_{\ell ms \ell'm'} (-1)^s 
        \ _u a_{\ell m} \ _u \hatb^{(j)*}_{\ell s}\,  \cpem{j,v} \ _{s+v}\tomega^{(j)}_{\ell'm'}
\onetwocols{\cols}{}{\nonumber \\ &\quad \times}
  \int \dd \vecr \ _{-s}Y_{\ell m}(\vecr) \ _{s+v} Y_{\ell'm'}(\vecr) \ _x
   Y^*_{\ell''m''}(\vecr)  \nonumber \\
  &= \sum_u \frac{k_{u}}{k_{v}} \sum_{j \ell m s \ell' m'} %e^{i v \gamma_{j}} 
        \ _u a_{\ell m} \ _u \hatb^{(j)*}_{\ell s}\,  \cpem{j,v} \ _{s+v}\tomega^{(j)}_{\ell'm'} 
         (-1)^{s+x+m''+\ell+\ell'+\ell''}
\onetwocols{\cols}{}{\nonumber \\ &\quad \times}
        \left[\frac{(2\ell+1)(2\ell'+1)(2\ell''+1)}{4\pi}\right]^{1/2}%%% (-1)^{\ell+\ell'+\ell''}
\onetwocols{\cols}{}{\nonumber \\ &\quad \times}
        \wjjj{\ell}{\ell'}{\ell''}{m}{m'}{-m''} \wjjj{\ell}{\ell'}{\ell''}{-s}{s+v}{-x}
\label{eq:map_SH}
\end{align}
which are only non-zero when $x=v$.
The cross power spectrum of spin $v_1$ and $v_2$ maps is then given by Eq.~(\ref{eq:crossCl_general}).

%------------------------------
\section{Hit matrix}
\label{appendix:hit_matrix}
%------------------------------

\newcommand{\za}{\ensuremath{z_2}}
\newcommand{\zb}{\ensuremath{z_4}}
\newcommand{\cza}{\ensuremath{\bar{z}_2}}
\newcommand{\czb}{\ensuremath{\bar{z}_4}}
\newcommand{\rhoa}{|\za|}
\newcommand{\rhob}{|\zb|}
\newcommand{\zab}{\za\czb}
\newcommand{\czab}{\cza\zb}

Introducing, for detector $j$,
\begin{equation}
\matH_{s}^{(j)} = 
\matrixthree%
{        {\omega_{s  }^{(j)}}}
{\cpem{j}  {\omega_{s-2}^{(j)}}} %e^{-2i\gamma_j}}}
{\cpem{j}  {\omega_{s+2}^{(j)}}} %e^{ 2i\gamma_j}}}%%%%
{\cpem{j}  {\omega_{s+2}^{(j)}}} %e^{ 2i\gamma_j}}}%
{\cpem{j}^2{\omega_{s  }^{(j)}}} %
{\cpem{j}^2{\omega_{s+4}^{(j)}}} %e^{ 4i\gamma_j}}}%%%
{\cpem{j}  {\omega_{s-2}^{(j)}}} %e^{-2i\gamma_j}}}%
{\cpem{j}^2{\omega_{s-4}^{(j)}}} %e^{-4i\gamma_j}}}%
{\cpem{j}^2{\omega_{s  }^{(j)}}} %%%
,
\end{equation}
the Hermitian hit matrix for a weighted combination of detectors is
\begin{align}
   \matH &\equiv \sum_j w_j \matH_{0}^{(j)}, \\
         &= h \, \matrixthree%
 {1}{\cza}{\za}%
 {\za}{x}{\zb}%
 {\cza}{\czb}{x},
\end{align}
with $h, x$ real and $\za, \zb$ complex numbers,
and has for inverse
\begin{equation}
  \matH^{-1} = \frac{1}{h\Delta}
  \matrixthree%
  {x^2-\rhob^2}{\zab-x\cza}{\czab-x\za}%
  {\czab-x\za}{x-\rhoa^2}{\za^2-\zb}%
  {\zab-x\cza}{\cza^2-\czb}{x-\rhoa^2},
\end{equation}
with 
\begin{equation}
  \Delta = x^2 - 2x\rhoa^2 - \rhob^2 + \za^2\czb+\cza^2\zb.
\end{equation}

In Eq.~(\ref{eq:define_tildeomega}) we defined
\begin{equation}
        \vectorthree%
        {      \tomega_{s}^{(j)}[0]}%
        {\cpem{j}\tomega_{s+2}^{(j)}[2] } %e^{ 2i\gamma_j}}%
        {\cpem{j}\tomega_{s-2}^{(j)}[-2]} %e^{-2i\gamma_j}}
        \equiv \matH^{-1} 
        \vectorthree%
        {       \omega_{s  }^{(j)}}%
        {\cpem{j} \omega_{s+2}^{(j)}} %e^{ 2i\gamma_j}}%
        {\cpem{j} \omega_{s-2}^{(j)}} %e^{-2i\gamma_j}}
\end{equation}
for any value of $s$, which provides
\begin{equation}
        \vectorthree%
        {        \tomega_{s}^{(j)}[0]}%
        {\cpem{j}\tomega_{s}^{(j)}[2]}%
        {\cpem{j}\tomega_{s}^{(j)}[-2]}
        = \frac{1}{h\Delta} 
        \vectorthree%
        {%
        (x^2-\rhob^2)\,          \omega_{s}^{(j)} + 
        (\zab-x\cza) \,\cpem{j}\,\omega_{s+2}^{(j)} %e^{ 2i\gamma_j} 
        \onetwocols{\cols}{}{\\} + 
        (\czab-x\za) \,\cpem{j}\,\omega_{s-2}^{(j)} %e^{-2i\gamma_j}
        }%
        {%
        (x-\rhoa^2) \,\cpem{j}\,\omega_{s}^{(j)} + 
        (\czab-x\za)\,        \,\omega_{s-2}^{(j)} %e^{-2i\gamma_j} 
        \onetwocols{\cols}{}{\\} + 
        (\za^2-\zb) \,\cpem{j}\,\omega_{s-4}^{(j)} %e^{-4i\gamma_j}
        }%
        {%
        (x-\rhoa^2)  \,\cpem{j}\,\omega_{s}^{(j)} +
        (\zab-x\cza) \,        \,\omega_{s+2}^{(j)} %e^{ 2i\gamma_j} 
        \onetwocols{\cols}{}{\\} + 
        (\cza^2-\czb)\,\cpem{j}\,\omega_{s+4}^{(j)} %e^{ 4i\gamma_j}
        },%
\end{equation}
so that $\tomega_{s}^{(j)}$ is of spin $s$, 
provided $z_2$ and $z_4$ are of spin 2 and 4 respectively.
Since $\omega_{s}^{(j)} = \omega_{-s}^{(j)*}$, we get
$\tomega_{s}^{(j)*}[2] = \tomega_{-s}^{(j)}[-2]$.

By definition,
\begin{equation}
        \matrixthree%
        {        {\tomega_{s  }^{(j)}[0]}}
        {\cpem{j}  {\tomega_{s-2}^{(j)}[0] }} %e^{-2i\gamma_j}}}
        {\cpem{j}  {\tomega_{s+2}^{(j)}[0] }} %e^{ 2i\gamma_j}}}%%%%
        {\cpem{j}  {\tomega_{s+2}^{(j)}[2] }} %e^{ 2i\gamma_j}}}%
        {\cpem{j}^2{\tomega_{s  }^{(j)}[2] }}  %
        {\cpem{j}^2{\tomega_{s+4}^{(j)}[2] }} %e^{ 4i\gamma_j}}}%%%
        {\cpem{j}  {\tomega_{s-2}^{(j)}[-2]}} %e^{-2i\gamma_j}}}%
        {\cpem{j}^2{\tomega_{s-4}^{(j)}[-2]}} %e^{-4i\gamma_j}}}%
        {\cpem{j}^2{\tomega_{s  }^{(j)}[-2]}} %%%
        =
        \matH^{-1}.
        \matH_{s}^{(j)}
\end{equation}
so that
\begin{equation}
        \sum_j w_j
        \matrixthree%
        {        {\tomega_{0  }^{(j)}[0]}}
        {\cpem{j}  {\tomega_{ -2}^{(j)}[0] }} %e^{-2i\gamma_j}}}
        {\cpem{j}  {\tomega_{  2}^{(j)}[0] }} %e^{ 2i\gamma_j}}}%%%%
        {\cpem{j}  {\tomega_{  2}^{(j)}[2] }} %e^{ 2i\gamma_j}}}%
        {\cpem{j}^2{\tomega_{0  }^{(j)}[2] }}%
        {\cpem{j}^2{\tomega_{  4}^{(j)}[2] }} %e^{ 4i\gamma_j}}}%%%
        {\cpem{j}  {\tomega_{ -2}^{(j)}[-2]}} %e^{-2i\gamma_j}}}%
        {\cpem{j}^2{\tomega_{ -4}^{(j)}[-2]}} %e^{-4i\gamma_j}}}%
        {\cpem{j}^2{\tomega_{0  }^{(j)}[-2]}}%%%
        =
        \matrixthree%
        {1}{0}{0}{0}{1}{0}{0}{0}{1}.
        \label{eq:omega0_identity}
\end{equation}
%

%------------------------------
\section{Wigner 3J symbols}
\label{appendix:3J}
%------------------------------
The Wigner 3J symbols describe the coupling between different spin weighted spherical harmonics at the same location:
\begin{align}
 _{s_1}Y_{\ell_1 m_1}(\vecr)\,
 _{s_2}Y_{\ell_2 m_2}(\vecr) & = \sum_{\ell_3 s_3 m_3}
 \left(\frac{(2\ell_1+1)(2\ell_2+1)(2\ell_3+1)}{4\pi}\right)^{1/2}
\onetwocols{\cols}{}{\nonumber \\ &\times}
 \wjjj{\ell_1}{\ell_2}{\ell_3}{m_1 }{m_2 }{m_3}
 \wjjj{\ell_1}{\ell_2}{\ell_3}{-s_1 }{-s_2 }{-s_3}\,
 _{s_3}Y^*_{\ell_3 m_3}(\vecr)
\label{eq:Ylm_product}
\end{align}
and the symbol
%\begin{align}
$\wjjj{\ell_1}{\ell_2}{\ell_3}{m_1 }{m_2 }{m_3}$
%\end{align}
is non-zero only when, $|m_i|\le \ell_i$ for $i=1,2,3$, $m_1+m_2+m_3=0$ and
\begin{align}
        |\ell_1-\ell_2|\le \ell_3 \le \ell_1+\ell_2.
        \label{eq:triangle_wigner}
\end{align}
They obey the relations
\begin{equation}
\wjjj{\ell_1}{\ell_2}{\ell_3}{-m_1 }{-m_2 }{-m_3} = (-1)^{\ell_1+\ell_2+\ell_3}
\wjjj{\ell_1}{\ell_2}{\ell_3}{m_1 }{m_2 }{m_3},
\end{equation}
and
\begin{equation}
\wjjj{\ell}{\ell}{0}{m}{-m}{0} = \frac{(-1)^{\ell-m}}{\sqrt{2\ell+1}}.
\end{equation}
Their standard orthogonality relations are
\begin{equation}
\sum_{\ell_3} (2 \ell_3 +1)\ \wjjj{\ell_1}{\ell_2}{\ell_3}{m_1 }{m_2 }{m_3}
                         \wjjj{\ell_1}{\ell_2}{\ell_3}{m_1'}{m_2'}{m_3'} = 
                        \delta_{m_1 m_1'} \delta_{m_2 m_2'},
\label{eq:w3j_ortho1}
\end{equation}
and
\begin{equation}
 \sum_{m_1m_2} \wjjj{\ell_1}{\ell_2}{\ell_3 }{m_1}{m_2}{m_3}
              \wjjj{\ell_1}{\ell_2}{\ell_3'}{m_1}{m_2}{m_3'} = 
                        \delta_{\ell_3 \ell_3'} \delta_{m_3 m_3'} \frac{\delta(\ell_1,\ell_2,\ell_3)}{2\ell_3+1},
\label{eq:w3j_ortho2}
\end{equation}
where $\delta(\ell_1,\ell_2,\ell_3)=1$ when $\ell_1,\ell_2,\ell_3$ obey the triangle relation
of Eq.~(\ref{eq:triangle_wigner}) and vanishes otherwise.

For $\ell_1 \ll \ell_2, \ell_3$ \citep[][Eq.~A2.1]{Edmonds1957}
\begin{equation}
\wjjj{\ell_1}{\ell_2}{\ell_3}{m_1 }{m_2}{-m_1-m_2} \simeq
\frac{(-1)^{\ell_3+m_2+m_1}}{\sqrt{2\ell_3+1}} d^{\ell_1}_{\ell_3-\ell_2,m_1} (\theta),
\end{equation}
where $d$ is the Wigner rotation matrix and $\cos \theta = 2 m_2 / (2\ell_2+1)$.
As a consequence, for $|m_2| \ll \ell_2$
\begin{equation}
\wjjj{\ell_1}{\ell_2}{\ell_3}{m_1 }{m_2}{-m_1-m_2} \simeq (-1)^{m_2-m_2'}
\wjjj{\ell_1}{\ell_2}{\ell_3}{m_1 }{m_2'}{-m_1-m_2'},
\label{eq:approx_sharp3j}
\end{equation}
and an approximate orthogonality relation can therefore be written, for $\ell_1,|m_1|,|m_2| \ll \ell_2, \ell_3$
\begin{equation}
\sum_{\ell_3} (2 \ell_3 +1)\ \wjjj{\ell_1}{\ell_2}{\ell_3}{m_1}{m_2}{m_3}
                         \wjjj{\ell_1}{\ell_2}{\ell_3}{m_1'}{m_2'}{m_3'} \simeq
                        (-1)^{m_2-m_2'} \delta_{m_1 m_1'}. 
\label{eq:w3j_ortho3}
\end{equation}

%------------------------------
\section{Spin weighted power spectra}
\label{appendix:cl_rot}
Since a complex field of spin $s$ can be written as $C_s = R_s + i I_s$ where $R_s$ and $I_s$ are real,
with
\begin{align}
R_s \pm i I_s = \sum_{\ell m} \ _{\pm s} a_{\ell m}\  _{\pm s} Y_{\ell m}
\end{align}
and, with the Condon-Shortley phase convention $_{s} Y^{*}_{\ell m} = (-1)^{s+m}\  _{-s} Y_{\ell -m},$
% \begin{align}
%       _{s} Y^{*}_{\ell m} &= (-1)^{s+m}\  _{-s} Y_{\ell -m},
% \end{align}
then
\begin{align}
        _{s}a^{*}_{\ell m} &= (-1)^{s+m} \ _{-s} a_{\ell-m}.
\end{align}
When $s=2$, one defines
\begin{subequations}
\begin{align}
        a^{E}_{\ell m} &= -\left( \ _{2}a_{\ell m}  + \ _{-2}a_{\ell m} \right)/2 \\
        a^{B}_{\ell m} &= -\left( \ _{2}a_{\ell m}  - \ _{-2}a_{\ell m} \right)/(2i)
\end{align}
\end{subequations}
such that $a^{X*}_{\ell m} = (-1)^{m} a^{X}_{\ell-m}$, with $X=E,B$,
% \begin{align}
%       a^{X*}_{\ell m} &= (-1)^{m} a^{X}_{\ell-m}, \quad {\rm with }\ X=E,B,
% \end{align}
and
\begin{subequations}
\begin{align}
        C^{EE}_{\ell} &=  \left(C_{\ell}^{22} + C_{\ell}^{2-2} + C_{\ell}^{-22} + C_{\ell}^{-2-2} \right) /4, \\
        C^{BB}_{\ell} &=  \left(C_{\ell}^{22} - C_{\ell}^{2-2} - C_{\ell}^{-22} + C_{\ell}^{-2-2} \right) /4, \\
%       C^{TE}_{\ell} &= -\left(C_{\ell}^{02} + C_{\ell}^{0-2} \right) /2 \\
%       C^{TB}_{\ell} &=  \left(C_{\ell}^{02} - C_{\ell}^{0-2} \right) /(2i) \\
        C^{EB}_{\ell} &= -\left(C_{\ell}^{22} - C_{\ell}^{2-2} + C_{\ell}^{-22} - C_{\ell}^{-2-2} \right) /(4i).
\end{align}
\end{subequations}
When $s=1$, one defines
\begin{subequations}
\begin{align}
        a^{G}_{\ell m} &= -\left( \ _{1}a_{\ell m}  - \ _{-1}a_{\ell m} \right)/2 \\
        a^{C}_{\ell m} &= -\left( \ _{1}a_{\ell m}  + \ _{-1}a_{\ell m} \right)/(2i)
\end{align}
\end{subequations}
such that $a^{X*}_{\ell m} = (-1)^{m} a^{X}_{\ell-m}$, with $X=G,C$,
% \begin{align}
%       a^{X*}_{\ell m} &= (-1)^{m} a^{X}_{\ell-m}, \quad {\rm with }\ X=G,C,
% \end{align}
and
\begin{subequations}
\begin{align}
        C^{GG}_{\ell} &=  \left(C_{\ell}^{11} - C_{\ell}^{1-1} - C_{\ell}^{-11} + C_{\ell}^{-1-1} \right) /4, \\
        C^{CC}_{\ell} &=  \left(C_{\ell}^{11} + C_{\ell}^{1-1} + C_{\ell}^{-11} + C_{\ell}^{-1-1} \right) /4, \\
%       C^{TG}_{\ell} &= -\left(C_{\ell}^{01} - C_{\ell}^{0-1} \right) /2 \\
%       C^{TC}_{\ell} &=  \left(C_{\ell}^{01} + C_{\ell}^{0-1} \right) /(2i) \\
        C^{GC}_{\ell} &= -\left(C_{\ell}^{11} + C_{\ell}^{1-1} - C_{\ell}^{-11} - C_{\ell}^{-1-1} \right) /(4i).
\end{align}
\end{subequations}

%---------------------------------------------------------------------
\section{Window matrices $W^{XY,\; X'Y'}_\ell$}
\label{appendix:window_matrices}

\subsection{Arbitrary beams, smooth scanning case}
\label{appendix:smooth_scanning}
Let us come back to Eqs.~(\ref{eq:cl_arbitrary}) and (\ref{eq:Omega_def}). 
These  can be cast in a more compact matrix form
\begin{align}
\widetilde{\matC}_{\ell}
& =
\sum_{j_1j_2}\sum_s
\left\{
\left[
\matD^{-1}.
\hat{\matB}^{(j_1)\dagger}_{\ell,s}.
\matD\ . 
%\onetwocols{\cols}{}{\right.\right. \nonumber \\ & \left.\left.} %% for 2 columns
% \matrixthree%
% {C_{\ell}^{00}}%
% {C_{\ell}^{02}}%
% {C_{\ell}^{0-2}}%
% {C_{\ell}^{20}}%
% {C_{\ell}^{22}}%
% {C_{\ell}^{2-2}}%
% {C_{\ell}^{-20}}%
% {C_{\ell}^{-22}}%
% {C_{\ell}^{-2-2}}%
\  \matC_{\ell} \ 
.\ 
\matD.
\hat{\matB}^{(j_2)}_{\ell,s}.
\matD^{-1}
\right]
*
\widetilde{\bf \Omega}^{(j_1j_2)}_s
 %e^{is(\gamma_{j_1}-\gamma_{j_2})}
\right\}
\label{eq:main}
\end{align}
where
\begin{align}
        \matC_{\ell} &\equiv
        \matrixthree%
        {C_{\ell}^{00}}%
        {C_{\ell}^{02}}%
        {C_{\ell}^{0-2}}%
        {C_{\ell}^{20}}%
        {C_{\ell}^{22}}%
        {C_{\ell}^{2-2}}%
        {C_{\ell}^{-20}}%
        {C_{\ell}^{-22}}%
        {C_{\ell}^{-2-2}},%
\end{align}
\begin{equation}
        \matD \equiv \matrixthree%
        {1}{0}{0}%
        {0}{1/2}{0}%
        {0}{0}{1/2},%
\end{equation}
\begin{equation}
\hat{\matB}^{(j)}_{\ell,s} \equiv
\matrixthree%
{ \ _{ 0}  \hatb^{(j)}_{\ell,s  }}%
{ \ _{ 0}  \hatb^{(j)}_{\ell,s-2}}%
{ \ _{ 0}  \hatb^{(j)}_{\ell,s+2}}%
{ \ _{ 2}  \hatb^{(j)}_{\ell,s  }}%
{ \ _{ 2}  \hatb^{(j)}_{\ell,s-2}}%
{ \ _{ 2}  \hatb^{(j)}_{\ell,s+2}}%
{ \ _{-2}  \hatb^{(j)}_{\ell,s  }}%
{ \ _{-2}  \hatb^{(j)}_{\ell,s-2}}%
{ \ _{-2}  \hatb^{(j)}_{\ell,s+2}}%
%, \\
% & =  &
=
\matrixthree%
{        \hatb^{(j)}_{\ell,s  }}%
{        \hatb^{(j)}_{\ell,s-2}}%
{        \hatb^{(j)}_{\ell,s+2}}%
{\cpei{j}\hatb^{(j)}_{\ell,s+2}}%
{\cpei{j}\hatb^{(j)}_{\ell,s  }}%
{\cpei{j}\hatb^{(j)}_{\ell,s+4}}%
{\cpei{j}\hatb^{(j)}_{\ell,s-2}}%
{\cpei{j}\hatb^{(j)}_{\ell,s-4}}%
{\cpei{j}\hatb^{(j)}_{\ell,s  }},%
\label{eq:beam_mat1}
\end{equation}
and $\matX*\matY$ denotes the elementwise product 
(also known as Hadamard or Schur product) of arrays $\matX$ and $\matY$.
Noting that 
%(\S\ref{appendix:alm_rot})
\begin{equation}
\matrixthree%
{C_{\ell}^{00}}%
{C_{\ell}^{02}}%
{C_{\ell}^{0-2}}%
{C_{\ell}^{20}}%
{C_{\ell}^{22}}%
{C_{\ell}^{2-2}}%
{C_{\ell}^{-20}}%
{C_{\ell}^{-22}}%
{C_{\ell}^{-2-2}}%
=
\matR_2.
\matrixthree%
{C_{\ell}^{TT}}%
{C_{\ell}^{TE}}%
{C_{\ell}^{TB}}%
{C_{\ell}^{ET}}%
{C_{\ell}^{EE}}%
{C_{\ell}^{EB}}%
{C_{\ell}^{BT}}%
{C_{\ell}^{BE}}%
{C_{\ell}^{BB}}%
.\matR_2^{\dagger}
\label{eq:Cl_stokes2spin}
\end{equation}
where $\matR_2$ was introduced in Eq.~(\ref{eq:rotation_matrix2_def_main}),
which leads to Eq.~(\ref{eq:beam_matrix}) that we recall here for convenience:
\begin{equation}
        \pC^{XY}_{\ell} = \sum_{X'Y'} W^{XY,\; X'Y'}_\ell C^{X'Y'}_{\ell}.
\end{equation}

Introducing the short-hand
\begin{equation}
        \pO_{v_1v_2} \equiv \tOmega_{v_1,v_2,s}^{(j_1j_2)}, % e^{is(\gamma_{j_1}-\gamma_{j_2})},
        \label{eq:def_Omega_hat}
\end{equation} 
describing the coupled moments of the polarized detectors $j_1$ and $j_2$
orientation, and assuming 
in Eq.~(\ref{eq:beam_mat1}) the beams to be perfectly 
co-polarized, with polar efficiencies $\cpei{j}$,
one gets, for $XY = TT, EE, BB, TE, TB, EB, ET, BT, BE$:
%
%\begin{widetext}
%%%%%%%%%% mathematica %%%%%%%%%
\begin{subequations}
\label{eq:wl_xyXY}
\begin{align} 
 W_{\ell}^{XY,\; TT} &= 
\sum_s \sum_{j_1j_2} 
 \left( 
\begin{array}{l}
% TT
 \pO_{00} \hatb_{\ell,s}^{(j_1)*} \hatb_{\ell,s}^{(j_2)}  \\
% EE
 \hatb_{\ell,s+2}^{(j_1)*} \left( \pO_{-2-2} \hatb_{\ell,s+2}^{(j_2)}+\pO_{-22} \hatb_{\ell,s-2}^{(j_2)} \right) 
+\hatb_{\ell,s-2}^{(j_1)*} \left( \pO_{2-2}  \hatb_{\ell,s+2}^{(j_2)}+\pO_{22}  \hatb_{\ell,s-2}^{(j_2)} \right)  \\
% BB
 \hatb_{\ell,s+2}^{(j_1)*} \left( \pO_{-2-2} \hatb_{\ell,s+2}^{(j_2)}-\pO_{-22} \hatb_{\ell,s-2}^{(j_2)} \right) 
+\hatb_{\ell,s-2}^{(j_1)*} \left( \pO_{22}   \hatb_{\ell,s-2}^{(j_2)}-\pO_{2-2} \hatb_{\ell,s+2}^{(j_2)} \right)  \\
% TE
 - \hatb_{\ell,s}^{(j_1)*} \left( \pO_{0-2} \hatb_{\ell,s+2}^{(j_2)}+\pO_{02} \hatb_{\ell,s-2}^{(j_2)} \right)  \\
% TB
 -i \hatb_{\ell,s}^{(j_1)*} \left( \pO_{02} \hatb_{\ell,s-2}^{(j_2)}-\pO_{0-2} \hatb_{\ell,s+2}^{(j_2)} \right)  \\
% EB
 i \hatb_{\ell,s+2}^{(j_1)*} \left( \pO_{-22} \hatb_{\ell,s-2}^{(j_2)}-\pO_{-2-2} \hatb_{\ell,s+2}^{(j_2)} \right) 
+i \hatb_{\ell,s-2}^{(j_1)*} \left( \pO_{22}  \hatb_{\ell,s-2}^{(j_2)}-\pO_{2-2}  \hatb_{\ell,s+2}^{(j_2)} \right) \\
% ET
 -\hatb_{\ell,s}^{(j_2)} \left(\pO_{-20} \hatb_{\ell,s+2}^{(j_1)*}+\pO_{20} \hatb_{\ell,s-2}^{(j_1)*}\right) \\
% BT
 -i \hatb_{\ell,s}^{(j_2)} \left(\pO_{-20} \hatb_{\ell,s+2}^{(j_1)*}-\pO_{20} \hatb_{\ell,s-2}^{(j_1)*}\right) \\
% BE
 i \hatb_{\ell,s+2}^{(j_1)*} \left( \pO_{-2-2} \hatb_{\ell,s+2}^{(j_2)}+\pO_{-22} \hatb_{\ell,s-2}^{(j_2)} \right) 
-i \hatb_{\ell,s-2}^{(j_1)*} \left( \pO_{2-2}  \hatb_{\ell,s+2}^{(j_2)}+\pO_{22}  \hatb_{\ell,s-2}^{(j_2)} \right)  \\
\end{array}
 \right),
 \label{eq:wl_xyTT}
\end{align}
which is illustrated in Fig.~\ref{fig:Wmatrix};
\begin{align} 
 W_{\ell}^{XY,\; EE} &= 
\sum_s \sum_{j_1j_2} 
\frac{\cpei{j_1} \cpei{j_2}}{4}
 \left( 
\begin{array}{l}
%1
 \pO_{00}   \left( \hatb_{\ell,s-2}^{(j_1)*}+\hatb_{\ell,s+2}^{(j_1)*}\right) 
            \left( \hatb_{\ell,s-2}^{(j_2)} +\hatb_{\ell,s+2}^{(j_2)} \right)\\
%2
 \hatb_{\ell,s}^{(j_1)*} \left[ \hatb_{\ell,s}^{(j_2)} \left( \pO_{-2-2}+\pO_{-22}+\pO_{2-2}+\pO_{22} \right) +\hatb_{\ell,s+4}^{(j_2)} \left( \pO_{-2-2}+\pO_{2-2} \right) +\hatb_{\ell,s-4}^{(j_2)} \left( \pO_{-22}+\pO_{22} \right)\right] \\ 
   \quad\quad +\hatb_{\ell,s+4}^{(j_1)*} \left[ \hatb_{\ell,s}^{(j_2)} \left( \pO_{-2-2}+\pO_{-22} \right) +\pO_{-2-2} \hatb_{\ell,s+4}^{(j_2)}+\pO_{-22} \hatb_{\ell,s-4}^{(j_2)} \right] \\
   \quad\quad +\hatb_{\ell,s-4}^{(j_1)*} \left[ \hatb_{\ell,s}^{(j_2)} \left( \pO_{2-2}+\pO_{22} \right) +\pO_{2-2} \hatb_{\ell,s+4}^{(j_2)}+\pO_{22} \hatb_{\ell,s-4}^{(j_2)} \right]  \\
%3
 \hatb_{\ell,s}^{(j_1)*} \left[ \hatb_{\ell,s}^{(j_2)} \left( \pO_{-2-2}-\pO_{-22}-\pO_{2-2}+\pO_{22} \right) +\hatb_{\ell,s+4}^{(j_2)} \left( \pO_{-2-2}-\pO_{2-2} \right) +\hatb_{\ell,s-4}^{(j_2)} \left( \pO_{22}-\pO_{-22} \right)\right] \\
   \quad\quad +\hatb_{\ell,s+4}^{(j_1)*} \left[ \hatb_{\ell,s}^{(j_2)} \left( \pO_{-2-2}-\pO_{-22} \right) +\pO_{-2-2} \hatb_{\ell,s+4}^{(j_2)}-\pO_{-22} \hatb_{\ell,s-4}^{(j_2)} \right] \\
   \quad\quad +\hatb_{\ell,s-4}^{(j_1)*} \left[ \hatb_{\ell,s}^{(j_2)} \left( \pO_{22}-\pO_{2-2} \right) -\pO_{2-2} \hatb_{\ell,s+4}^{(j_2)}+\pO_{22} \hatb_{\ell,s-4}^{(j_2)} \right]  \\
%4
 - \left(\hatb_{\ell,s-2}^{(j_1)*}+\hatb_{\ell,s+2}^{(j_1)*}\right) \left[ \hatb_{\ell,s}^{(j_2)} \left( \pO_{0-2}+\pO_{02} \right) +\pO_{0-2} \hatb_{\ell,s+4}^{(j_2)}+\pO_{02} \hatb_{\ell,s-4}^{(j_2)} \right]  \\
%5
 -i \left(\hatb_{\ell,s-2}^{(j_1)*}+\hatb_{\ell,s+2}^{(j_1)*}\right) \left[ \hatb_{\ell,s}^{(j_2)} \left( \pO_{02}-\pO_{0-2} \right) -\pO_{0-2} \hatb_{\ell,s+4}^{(j_2)}+\pO_{02} \hatb_{\ell,s-4}^{(j_2)} \right]  \\
%6
 i \hatb_{\ell,s}^{(j_1)*} \left[ \hatb_{\ell,s}^{(j_2)} \left( -\pO_{-2-2}+\pO_{-22}-\pO_{2-2}+\pO_{22} \right) -\hatb_{\ell,s+4}^{(j_2)} \left( \pO_{-2-2}+\pO_{2-2} \right) +\hatb_{\ell,s-4}^{(j_2)} \left( \pO_{-22}+\pO_{22} \right)\right] \\ 
   \quad\quad+i\hatb_{\ell,s+4}^{(j_1)*} \left[ \hatb_{\ell,s}^{(j_2)} \left( \pO_{-22}-\pO_{-2-2} \right) -\pO_{-2-2} \hatb_{\ell,s+4}^{(j_2)}+\pO_{-22} \hatb_{\ell,s-4}^{(j_2)} \right] \\
   \quad\quad+i\hatb_{\ell,s-4}^{(j_1)*} \left[ \hatb_{\ell,s}^{(j_2)} \left( \pO_{22}-\pO_{2-2} \right) -\pO_{2-2} \hatb_{\ell,s+4}^{(j_2)}+\pO_{22} \hatb_{\ell,s-4}^{(j_2)} \right] \\
%7
 -\left( \hatb_{\ell,s-2}^{(j_2)}+\hatb_{\ell,s+2}^{(j_2)} \right)  \left[\left(\pO_{-20}+\pO_{20} \right)  \hatb_{\ell,s}^{(j_1)*}+\pO_{-20} \hatb_{\ell,s+4}^{(j_1)*}+\pO_{20} \hatb_{\ell,s-4}^{(j_1)*}\right] \\
%8
 i \left( \hatb_{\ell,s-2}^{(j_2)}+\hatb_{\ell,s+2}^{(j_2)} \right)  \left[\left(\pO_{20}-\pO_{-20} \right)  \hatb_{\ell,s}^{(j_1)*}-\pO_{-20} \hatb_{\ell,s+4}^{(j_1)*}+\pO_{20} \hatb_{\ell,s-4}^{(j_1)*}\right] \\
%9
 i \hatb_{\ell,s}^{(j_1)*} \left[ \hatb_{\ell,s}^{(j_2)} \left( \pO_{-2-2}+\pO_{-22}-\pO_{2-2}-\pO_{22} \right) +\hatb_{\ell,s+4}^{(j_2)} \left( \pO_{-2-2}-\pO_{2-2} \right) +\hatb_{\ell,s-4}^{(j_2)} \left( \pO_{-22}-\pO_{22} \right)\right] \\ 
   \quad\quad +i\hatb_{\ell,s+4}^{(j_1)*} \left[ \hatb_{\ell,s}^{(j_2)} \left( \pO_{-2-2}+\pO_{-22} \right) +\pO_{-2-2} \hatb_{\ell,s+4}^{(j_2)}+\pO_{-22} \hatb_{\ell,s-4}^{(j_2)} \right] \\
   \quad\quad -i\hatb_{\ell,s-4}^{(j_1)*} \left[ \hatb_{\ell,s}^{(j_2)} \left( \pO_{2-2}+\pO_{22} \right) +\pO_{2-2} \hatb_{\ell,s+4}^{(j_2)}+\pO_{22} \hatb_{\ell,s-4}^{(j_2)} \right] \\
\end{array}
 \right),
 \label{eq:wl_xyEE}
\end{align}
\begin{align} 
 W_{\ell}^{XY,\; TE} &= 
\sum_s \sum_{j_1j_2} 
\frac{\cpei{j_2}}{2}
 \left( 
\begin{array}{l}
%1
 -\pO_{00} \hatb_{\ell,s}^{(j_1)*} \left( \hatb_{\ell,s-2}^{(j_2)}+\hatb_{\ell,s+2}^{(j_2)} \right)   \\
%2
 -\hatb_{\ell,s+2}^{(j_1)*} \left[ \hatb_{\ell,s}^{(j_2)} \left( \pO_{-2-2}+\pO_{-22} \right) +\pO_{-2-2} \hatb_{\ell,s+4}^{(j_2)}+\pO_{-22} \hatb_{\ell,s-4}^{(j_2)} \right] \\
  \quad\quad -\hatb_{\ell,s-2}^{(j_1)*} \left[ \hatb_{\ell,s}^{(j_2)} \left( \pO_{2-2}+\pO_{22} \right) +\pO_{2-2} \hatb_{\ell,s+4}^{(j_2)}+\pO_{22} \hatb_{\ell,s-4}^{(j_2)} \right]  \\
%3
 -\hatb_{\ell,s+2}^{(j_1)*} \left[ \hatb_{\ell,s}^{(j_2)} \left( \pO_{-2-2}-\pO_{-22} \right) +\pO_{-2-2} \hatb_{\ell,s+4}^{(j_2)}-\pO_{-22} \hatb_{\ell,s-4}^{(j_2)} \right] \\
  \quad\quad -\hatb_{\ell,s-2}^{(j_1)*} \left[ \hatb_{\ell,s}^{(j_2)} \left( \pO_{22}-\pO_{2-2} \right) -\pO_{2-2} \hatb_{\ell,s+4}^{(j_2)}+\pO_{22} \hatb_{\ell,s-4}^{(j_2)} \right]  \\
%4
 \hatb_{\ell,s}^{(j_1)*} \left[ \hatb_{\ell,s}^{(j_2)} \left( \pO_{0-2}+\pO_{02} \right) +\pO_{0-2} \hatb_{\ell,s+4}^{(j_2)}+\pO_{02} \hatb_{\ell,s-4}^{(j_2)} \right]  \\
%5
 i \hatb_{\ell,s}^{(j_1)*} \left[ \hatb_{\ell,s}^{(j_2)} \left( \pO_{02}-\pO_{0-2} \right) -\pO_{0-2} \hatb_{\ell,s+4}^{(j_2)}+\pO_{02} \hatb_{\ell,s-4}^{(j_2)} \right]  \\
%6
 -i \hatb_{\ell,s+2}^{(j_1)*} \left[ \hatb_{\ell,s}^{(j_2)} \left( \pO_{-22}-\pO_{-2-2} \right) -\pO_{-2-2} \hatb_{\ell,s+4}^{(j_2)}+\pO_{-22} \hatb_{\ell,s-4}^{(j_2)} \right] \\
  \quad\quad -i\hatb_{\ell,s-2}^{(j_1)*} \left[ \hatb_{\ell,s}^{(j_2)} \left( \pO_{22}-\pO_{2-2} \right) -\pO_{2-2} \hatb_{\ell,s+4}^{(j_2)}+\pO_{22} \hatb_{\ell,s-4}^{(j_2)} \right] \\
%7
 \left( \hatb_{\ell,s-2}^{(j_2)}+\hatb_{\ell,s+2}^{(j_2)} \right)  \left(\pO_{-20} \hatb_{\ell,s+2}^{(j_1)*}+\pO_{20} \hatb_{\ell,s-2}^{(j_1)*}\right) \\
%8
 -i \left( \hatb_{\ell,s-2}^{(j_2)}+\hatb_{\ell,s+2}^{(j_2)} \right)  \left(\pO_{20} \hatb_{\ell,s-2}^{(j_1)*}-\pO_{-20} \hatb_{\ell,s+2}^{(j_1)*}\right) \\
%9
 -i \hatb_{\ell,s+2}^{(j_1)*} \left[ \hatb_{\ell,s}^{(j_2)} \left( \pO_{-2-2}+\pO_{-22} \right) +\pO_{-2-2} \hatb_{\ell,s+4}^{(j_2)}+\pO_{-22} \hatb_{\ell,s-4}^{(j_2)} \right] \\
  \quad\quad +i \hatb_{\ell,s-2}^{(j_1)*} \left[ \hatb_{\ell,s}^{(j_2)} \left( \pO_{2-2}+\pO_{22} \right) +\pO_{2-2} \hatb_{\ell,s+4}^{(j_2)}+\pO_{22} \hatb_{\ell,s-4}^{(j_2)} \right]  \\
\end{array}
 \right).
 \label{eq:wl_xyTE}
\end{align}
\end{subequations}
%\end{widetext}
Since, by definition (Eqs.~\ref{eq:def_hatb} and \ref{eq:def_Omega_hat}),
$\hatb_{\ell,s}^{(j)*} = (-1)^s \hatb_{\ell,-s}^{(j)}$ and ${\pOs_{v_1,v_2}} =
\pOn_{-v_1,-v_2}$, one can check that each term of $W^{XY,\; X'Y'}_\ell$ is
real, as expected.\\

\subsection{Arbitrary beams, ideal scanning}
\label{appendix:ideal_scanning}
In the case of ideal scanning described in Section \ref{sec:ideal_scanning},
one gets $\omega^{(j)}_s(p) = \delta_{s,0} h(p)$, so that the hit matrix is diagonal:
\begin{align}
\matH(p) &= h(p)\sum_j\matrixthree
        {w_j}{0}{0}
        {0}{w_j\cpem{j}^2}{0}
        {0}{0}{w_j\cpem{j}^2},
\end{align}
and the orientation moments
\begin{align}
\tomega^{(j)}_s[0]    &= \delta_{s,0} \left(\sum_k w_k         \right)^{-1},
&
\tomega^{(j)}_s[\pm2] &= \delta_{s,0} \left(\sum_k w_k\cpem{k}^2\right)^{-1},
\end{align}
are such that
\begin{equation}
        \pO_{v_1v_2} =
\matrixthree%
{                    \myxi_{00}}%
{\cpem{j_2}          \myxi_{02}}%
{\cpem{j_2}          \myxi_{02}}%
{\cpem{j_1}          \myxi_{20}}%
{\cpem{j_1}\cpem{j_2}\myxi_{22}}%
{\cpem{j_1}\cpem{j_2}\myxi_{22}}%
{\cpem{j_1}          \myxi_{20}}%
{\cpem{j_1}\cpem{j_2}\myxi_{22}}%
{\cpem{j_1}\cpem{j_2}\myxi_{22}}%
\ 
\delta_{s,0},
\end{equation}
with
\begin{align}
\myxi_{00}^{-1} &= \sum_{k_1k_2}w_{k_1}w_{k_2}                        ,
&              
\myxi_{02}^{-1} &= \sum_{k_1k_2}w_{k_1}w_{k_2}            \cpem{k_2}^2,% \nonumber\\
&
\myxi_{20}^{-1} &= \sum_{k_1k_2}w_{k_1}w_{k_2}\cpem{k_1}^2            , 
&              
\myxi_{22}^{-1} &= \sum_{k_1k_2}w_{k_1}w_{k_2}\cpem{k_1}^2\cpem{k_2}^2.
\end{align}
One then obtains the beam matrices
%\begin{widetext}
%%%%%%%%%%%%%  mathematica
\begin{subequations}
 \label{eq:wl_xyXY_is}
\begin{align} 
 W_{\ell}^{XY,\; TT} &= 
\sum_{j_1j_2} 
 \left( 
\begin{array}{l}
%1 ----
 \hatb_{\ell,0}^{(j_2)} \hatb_{\ell,0}^{(j_1)*} 
{\ \myxi_{00}}   \\
%2
 \left( \hatb_{\ell,-2}^{(j_2)}+\hatb_{\ell,2}^{(j_2)} \right)  \left(\hatb_{\ell,-2}^{(j_1)*}+\hatb_{\ell,2}^{(j_1)*}\right) 
{\ \cpem{j_1}\cpem{j_2}\myxi_{22}}  \\
%3
 \left( \hatb_{\ell,-2}^{(j_2)}-\hatb_{\ell,2}^{(j_2)} \right)  \left(\hatb_{\ell,-2}^{(j_1)*}-\hatb_{\ell,2}^{(j_1)*}\right) 
{\ \cpem{j_1}\cpem{j_2}\myxi_{22}} \\
%4 ----
 -\left( \hatb_{\ell,-2}^{(j_2)}+\hatb_{\ell,2}^{(j_2)} \right)  \hatb_{\ell,0}^{(j_1)*} 
{\ \cpem{j_2}\myxi_{02}} \\
%5
 -i \left( \hatb_{\ell,-2}^{(j_2)}-\hatb_{\ell,2}^{(j_2)} \right)  \hatb_{\ell,0}^{(j_1)*} 
{\ \cpem{j_2}\myxi_{02}} \\
%6
 i \left( \hatb_{\ell,-2}^{(j_2)}-\hatb_{\ell,2}^{(j_2)} \right)  \left(\hatb_{\ell,-2}^{(j_1)*}+\hatb_{\ell,2}^{(j_1)*}\right) 
{\ \cpem{j_1}\cpem{j_2}\myxi_{22}} \\
%7 ----
 -\hatb_{\ell,0}^{(j_2)} \left(\hatb_{\ell,-2}^{(j_1)*}+\hatb_{\ell,2}^{(j_1)*}\right) 
{\ \cpem{j_1}\myxi_{20}} \\
%8
 i \hatb_{\ell,0}^{(j_2)} \left(\hatb_{\ell,-2}^{(j_1)*}-\hatb_{\ell,2}^{(j_1)*}\right) 
{\ \cpem{j_1}\myxi_{20}} \\
%9
 -i \left( \hatb_{\ell,-2}^{(j_2)}+\hatb_{\ell,2}^{(j_2)} \right)  \left(\hatb_{\ell,-2}^{(j_1)*}-\hatb_{\ell,2}^{(j_1)*}\right) 
{\ \cpem{j_1}\cpem{j_2}\myxi_{22}} \\
\end{array}
 \right),
 \label{eq:wl_xyTT_is}
\end{align}
\begin{align} 
 W_{\ell}^{XY,\; EE} &= 
\sum_{j_1j_2} 
\frac{\cpei{j_1} \cpei{j_2}}{4}
 \left( 
\begin{array}{l}
%1 ----
 \left( \hatb_{\ell,-2}^{(j_2)}+\hatb_{\ell,2}^{(j_2)} \right)  \left(\hatb_{\ell,-2}^{(j_1)*}+\hatb_{\ell,2}^{(j_1)*}\right) 
{\ \myxi_{00}}   \\
%2
 \left( \hatb_{\ell,-4}^{(j_2)}+2 \hatb_{\ell,0}^{(j_2)}+\hatb_{\ell,4}^{(j_2)} \right)  \left(\hatb_{\ell,-4}^{(j_1)*}+2 \hatb_{\ell,0}^{(j_1)*}+\hatb_{\ell,4}^{(j_1)*}\right) 
{\ \cpem{j_1}\cpem{j_2}\myxi_{22}}  \\
%3
 \left( \hatb_{\ell,-4}^{(j_2)}-\hatb_{\ell,4}^{(j_2)} \right)  \left(\hatb_{\ell,-4}^{(j_1)*}-\hatb_{\ell,4}^{(j_1)*}\right) 
{\ \cpem{j_1}\cpem{j_2}\myxi_{22}}  \\
%4 ----
 -\left( \hatb_{\ell,-4}^{(j_2)}+2 \hatb_{\ell,0}^{(j_2)}+\hatb_{\ell,4}^{(j_2)} \right)  \left(\hatb_{\ell,-2}^{(j_1)*}+\hatb_{\ell,2}^{(j_1)*}\right) 
{\ \cpem{j_2}\myxi_{02}}  \\
%5
 -i \left( \hatb_{\ell,-4}^{(j_2)}-\hatb_{\ell,4}^{(j_2)} \right)  \left(\hatb_{\ell,-2}^{(j_1)*}+\hatb_{\ell,2}^{(j_1)*}\right) 
{\ \cpem{j_2}\myxi_{02}}  \\
%6
 i \left( \hatb_{\ell,-4}^{(j_2)}-\hatb_{\ell,4}^{(j_2)} \right)  \left(\hatb_{\ell,-4}^{(j_1)*}+2 \hatb_{\ell,0}^{(j_1)*}+\hatb_{\ell,4}^{(j_1)*}\right) 
{\ \cpem{j_1}\cpem{j_2}\myxi_{22}}  \\
%7 ----
 -\left( \hatb_{\ell,-2}^{(j_2)}+\hatb_{\ell,2}^{(j_2)} \right)  \left(\hatb_{\ell,-4}^{(j_1)*}+2 \hatb_{\ell,0}^{(j_1)*}+\hatb_{\ell,4}^{(j_1)*}\right) 
{\ \cpem{j_1}\myxi_{20}}  \\
%8
 i \left( \hatb_{\ell,-2}^{(j_2)}+\hatb_{\ell,2}^{(j_2)} \right)  \left(\hatb_{\ell,-4}^{(j_1)*}-\hatb_{\ell,4}^{(j_1)*}\right) 
{\ \cpem{j_1}\myxi_{20}}  \\
%9
 -i \left( \hatb_{\ell,-4}^{(j_2)}+2 \hatb_{\ell,0}^{(j_2)}+\hatb_{\ell,4}^{(j_2)} \right)  \left(\hatb_{\ell,-4}^{(j_1)*}-\hatb_{\ell,4}^{(j_1)*}\right) 
{\ \cpem{j_1}\cpem{j_2}\myxi_{22}}  \\
\end{array}
 \right),
 \label{eq:wl_xyEE_is}
\end{align}
\begin{align} 
 W_{\ell}^{XY,\; TE} &= 
\sum_{j_1j_2} 
\frac{\cpei{j_2}}{2}
 \left( 
\begin{array}{l}
%1 ----
 -\left( \hatb_{\ell,-2}^{(j_2)}+\hatb_{\ell,2}^{(j_2)} \right)  \hatb_{\ell,0}^{(j_1)*} 
{\ \myxi_{00}}   \\
%2
 -\left( \hatb_{\ell,-4}^{(j_2)}+2 \hatb_{\ell,0}^{(j_2)}+\hatb_{\ell,4}^{(j_2)} \right)  \left(\hatb_{\ell,-2}^{(j_1)*}+\hatb_{\ell,2}^{(j_1)*}\right)
{\ \cpem{j_1}\cpem{j_2}\myxi_{22}}   \\
%3
 -\left( \hatb_{\ell,-4}^{(j_2)}-\hatb_{\ell,4}^{(j_2)} \right)  \left(\hatb_{\ell,-2}^{(j_1)*}-\hatb_{\ell,2}^{(j_1)*}\right)
{\ \cpem{j_1}\cpem{j_2}\myxi_{22}}   \\
%4 ----
 \left( \hatb_{\ell,-4}^{(j_2)}+2 \hatb_{\ell,0}^{(j_2)}+\hatb_{\ell,4}^{(j_2)} \right)  \hatb_{\ell,0}^{(j_1)*} 
{\ \cpem{j_2}\myxi_{02}}  \\
%5
 i \left( \hatb_{\ell,-4}^{(j_2)}-\hatb_{\ell,4}^{(j_2)} \right)  \hatb_{\ell,0}^{(j_1)*} 
{\ \cpem{j_2}\myxi_{02}}  \\
%6
 -i \left( \hatb_{\ell,-4}^{(j_2)}-\hatb_{\ell,4}^{(j_2)} \right)  \left(\hatb_{\ell,-2}^{(j_1)*}+\hatb_{\ell,2}^{(j_1)*}\right) 
{\ \cpem{j_1}\cpem{j_2}\myxi_{22}}  \\
%7 ----
 \left( \hatb_{\ell,-2}^{(j_2)}+\hatb_{\ell,2}^{(j_2)} \right)  \left(\hatb_{\ell,-2}^{(j_1)*}+\hatb_{\ell,2}^{(j_1)*}\right) 
{\ \cpem{j_1}\myxi_{20}}  \\
%8
 -i \left( \hatb_{\ell,-2}^{(j_2)}+\hatb_{\ell,2}^{(j_2)} \right)  \left(\hatb_{\ell,-2}^{(j_1)*}-\hatb_{\ell,2}^{(j_1)*}\right) 
{\ \cpem{j_1}\myxi_{20}}  \\
%9
 i \left( \hatb_{\ell,-4}^{(j_2)}+2 \hatb_{\ell,0}^{(j_2)}+\hatb_{\ell,4}^{(j_2)} \right)  \left(\hatb_{\ell,-2}^{(j_1)*}-\hatb_{\ell,2}^{(j_1)*}\right) 
{\ \cpem{j_1}\cpem{j_2}\myxi_{22}}  \\
\end{array}
 \right).
 \label{eq:wl_xyTE_is}
\end{align}
\end{subequations}
%\end{widetext}

The implications of Eq.~(\ref{eq:wl_xyXY_is}) are discussed in Section \ref{sec:ideal_scanning}.

%------------------------------
%\begin{widetext}
%------------------------------
\section{Finite pixel size}
\label{sec:append_subpixel}
Introducing the spin raising and lowering differential operators applied to a function $f$ of spin $s$,
\citep[][and references therein]{ZaldarriagaSeljak1997, Bunn+2003}
\begin{align}
\dbar f &= - \sin^{s} \theta
\left(\frac{\partial}{\partial \theta} + \frac{i}{\sin\theta}\frac{\partial}{\partial \varphi} \right)
\left[ \sin^{-s}\theta\ f \right]  
   \quad = s \cot\theta\ f - \frac{\partial f}{\partial \theta} - \frac{i}{\sin\theta}\frac{\partial f}{\partial \varphi}
\\
\bardbar f &= - \sin^{-s} \theta
\left(\frac{\partial}{\partial \theta} - \frac{i}{\sin\theta}\frac{\partial}{\partial \varphi} \right)
\left[ \sin^{s}\theta\ f \right] 
   \quad = -s \cot\theta\ f - \frac{\partial f}{\partial \theta} + \frac{i}{\sin\theta}\frac{\partial f}{\partial \varphi}
\end{align}
the spin weighed spherical harmonics are defined as
\begin{align}
_s Y_{\ell m} &\equiv \sqrt{\frac{(\ell-s)!}{(\ell+s)!}}\ \dbar^{s} Y_{\ell m}, \quad 0\le s \le \ell;\\
_s Y_{\ell m} &\equiv (-1)^s \sqrt{\frac{(\ell+s)!}{(\ell-s)!}}\ \bardbar^{-s} Y_{\ell m}, \quad -\ell\le s \le 0;
\end{align}
such that
\begin{align}
             \dbar \ _sY_{\ell m} &=  f(\ell, s)\ _{s+1}Y_{\ell m},\\
           \bardbar\ _sY_{\ell m} &= -f(\ell,-s)\ _{s-1}Y_{\ell m}, %\\
%       \bardbar \dbar \ _sY_{\ell m} & = -f(\ell,s)^2\ _{s}Y_{\ell m}, \\
%       \dbar \bardbar \ _sY_{\ell m} & = -f(\ell,-s)^2\ _{s}Y_{\ell m}, 
\end{align}
with $f(\ell,s) = \sqrt{(\ell-s)(\ell+s+1)} = \sqrt{\ell(\ell+1) - s(s+1)}$.

As noticed in \citet{Planck2013-7} and \citet{Planck2013-17}, the formalism of 
subpixel effect is very close to the one
of gravitational lensing described in \citet{Hu2000} and \citet{LewisChallinor2006}.

For $\vecr = (1,\theta,\varphi) = \vece_r$ and $\dd\vecr = (0,\dd\theta,\dd\varphi) = \dd\theta\, \vece_\theta + \sin \theta \dd\varphi\, \vece_\varphi$,
\begin{align}
_{s}Y_{\ell m}(\vecr+\dd\vecr) & =\ _{s}Y_{\ell m}(\vecr) + \dd\vecr . \nabla \ _{s}Y_{\ell m}(\vecr) + \frac{1}{2} \sum_{ij} \dd\vecr_i\dd\vecr_j  \nabla_i\nabla_j \ _{s}Y_{\ell m}(\vecr)\\
        & = \ _{s}Y_{\ell m}(\vecr) - \frac{1}{2}\left(\bardr\ \dbar 
+ \dr\ \bardbar\right)\ _{s}Y_{\ell m}(\vecr) + \frac{1}{8}\left(
\bardr\bardr\ \dbar\dbar +
\bardr\dr\ \dbar\bardbar +
\dr\bardr\ \bardbar\dbar +
\dr\dr\ \bardbar\bardbar \right)\ _{s}Y_{\ell m}(\vecr)\\
        & = \ _{s}Y_{\ell m}(\vecr) 
-\frac{1}{2}\left(\bardr f(\ell,s)\ _{s+1}Y_{\ell m}(\vecr)- \dr f(\ell,-s)\ _{s-1}Y_{\ell m}(\vecr)\right)
-\frac{1}{4}\dr\bardr \left(\ell(\ell+1) - s^2\right) \ _{s}Y_{\ell m}(\vecr) \nonumber \\
& \quad +\frac{1}{8}\left(\bardr\bardr\ g(\ell,s)\ _{s+2}Y_{\ell m}(\vecr)+
\dr\dr\ g(\ell,-s)\ _{s-2}Y_{\ell m}(\vecr)\right)
\end{align}
with $\dr = \dd\vecr . (\vece_\theta + i \vece_\varphi) = \dd\theta + i \sin\theta \dd\varphi$
, $\bardr = \dd\theta - i \sin\theta \dd\varphi$ and $g(\ell,s) = f(\ell,s)f(\ell,s+1)$.

Identifying $\dd\vecr$ to the position of a measurement relative to the nominal center $\vecr$ of the pixel to which it is attributed, this expansion of $_{s}Y_{\ell m}$ can be injected into Eqs.~(\ref{eq:TOD_from_alm}) and (\ref{eq:map_SH}).
Assuming $\dd\vecr$ to be uncorrelated with
the orientation of the detector, two extra terms, both quadratic in $\dd\vecr,$ will appear in the final power
spectra.

The first term involves the scalar product of the gradient of the signal in the pixel, assumed to be totally dominated by the temperature, with the weighted sum of $\dd\vecr$ over all samples in that pixel. Introducing
\begin{align}
        \cpem{j,v} \tomega_{s+v}^{(j)\pm}(p) = \sum_{v'} (\matH^{-1}(p))_{vv'} \cpem{j,v'} \sum_{t \in p} 
        (\dd\theta_t \pm i \sin\theta_t \dd\varphi_t) f_{j,t} e^{i (s+v') \alpha^{(j)}_t}
\end{align}
which is of spin $s+v\pm1$ and such that $\left(\cpem{j,v} \tomega_{s+v}^{(j)+}\right)^{*} = \cpem{j,-v} \tomega_{-s-v}^{(j)-}$, one finds
\begin{align}
\Delta \pC^{v_1v_2}_{\ell''} 
        &= 
        \frac{1}{k_{v_1}k_{v_2}} 
        \sum_{s_1 s_2} (-1)^{s_1+s_2} \sum_{j_1 j_2 \ell}
        \ell(\ell+1)
        \frac{2\ell+1}{4\pi} 
        C^{TT}_{\ell}
        \hatb^{(j_1)*}_{\ell s_1} \, \hatb^{(j_2)}_{\ell s_2} 
        \nonumber \\ &\ \times 
        \sum_{\ell'} 
        \frac{2\ell'+1}{4}\left[
          D^{(j_1j_2)++}_{s_1+v_1,s_2+v_2, \ell'}J^{v1,v2}_{s_1+1,s_2+1}
        + D^{(j_1j_2)--}_{s_1+v_1,s_2+v_2, \ell'}J^{v1,v2}_{s_1-1,s_2-1}
        - D^{(j_1j_2)+-}_{s_1+v_1,s_2+v_2, \ell'}J^{v1,v2}_{s_1+1,s_2-1}
        - D^{(j_1j_2)-+}_{s_1+v_1,s_2+v_2, \ell'}J^{v1,v2}_{s_1-1,s_2+1}
        \right]
        \label{eq:subpix_noise_general}
\end{align}
with
\begin{align}
        D^{(j_1j_2)\sigma_1\sigma_2}_{s_1+v_1,s_2+v_2, \ell'} &= 
          \cpem{j_1,v_1}\cpem{j_2,v_2}\ 
          \frac{1}{2\ell'+1}\sum_{m'}
          {}_{s_1+v_1}\tomega^{(j_1)\sigma_1 }_{\ell'm'}
          {}_{s_2+v_2}\tomega^{(j_2)\sigma_2*}_{\ell'm'} \quad \rm{with}\ \{\sigma_1,\sigma_2\} \in \{+,-\},\\
        J^{v1,v2}_{s_1+\sigma_1,s_2+\sigma_2} &=
        \wjjj{\ell}{\ell'}{\ell''}{-s_1-\sigma_1}{s_1+\sigma_1+v_1}{-v_1}
        \wjjj{\ell}{\ell'}{\ell''}{-s_2-\sigma_2}{s_2+\sigma_2+v_2}{-v_2} \quad \rm{with}\ \{\sigma_1,\sigma_2\} \in \{+1,-1\}.
\end{align}
In the case of temperature, and assuming the beams to be circular, this simplifies to
\begin{align}
\Delta \pC^{TT}_{\ell''} 
        &= 
        \sum_{j_1 j_2 \ell}
        \ell(\ell+1)
        \frac{2\ell+1}{4\pi} 
        C^{TT}_{\ell}
        \hatb^{(j_1)*}_{\ell 0} \, \hatb^{(j_2)}_{\ell 0} 
        \nonumber \\
        &\ \times 
        \sum_{\ell'} 
        \frac{2\ell'+1}{4}\wjjj{\ell}{\ell'}{\ell''}{1}{-1}{0}^2 \left[
          \left(D^{(j_1j_2)++}_{00, \ell'}
        + D^{(j_1j_2)--}_{00, \ell'}\right)
        - (-1)^{\ell+\ell'+\ell''}\left(D^{(j_1j_2)+-}_{00, \ell'}
        + D^{(j_1j_2)-+}_{00, \ell'}\right)
        \right]
\end{align}
in agreement with \citet{Planck2013-7}, once one identifies 
$\left(D^{(j_1j_2)++}_{00, \ell'} + D^{(j_1j_2)--}_{00, \ell'}\right)/2$ 
as the sum of the gradient and curl parts of the (spin 1) displacement field power spectrum and 
$-\left(D^{(j_1j_2)+-}_{00, \ell'} + D^{(j_1j_2)-+}_{00, \ell'}\right)/2$ 
as their difference (see Section \ref{appendix:cl_rot}).
\\
It is instructive to further assume the relative location of the hit's center of mass 
to be only weakly correlated between pixels,
so that all its derived power spectra can be assumed to be white:
$D^{(j_1j_2)\sigma_1\sigma_2}_{s_1+v_1,s_2+v_2, \ell'} = D^{(j_1j_2)\sigma_1\sigma_2}_{s_1+v_1,s_2+v_2}$
(i.e., with a variance  $D^{(j_1j_2)\sigma_1\sigma_2}_{s_1+v_1,s_2+v_2} / \Opix$ 
in pixels of solid angle $\Opix = 4 \pi/\npix$).
Equation~(\ref{eq:w3j_ortho1}) then ensures that the sub-pixel noise of 
Eq.~(\ref{eq:subpix_noise_general}) is also white, with constant polarized spectra
\begin{equation}
\Delta \pC^{XY}_{\ell''} = N^{XY} =  \sum_{\ell}
        \ell(\ell+1)
        \frac{2\ell+1}{4\pi} 
        C^{TT}_{\ell}
        {\cal W}^{XY}_\ell
\end{equation}
with    
\begin{align}
{\cal W}^{TT}_\ell
        & =\frac{1}{4}
        \sum_{j_1 j_2 s}\left(
        \hatb^{(j_1)*}_{\ell s} \, \hatb^{(j_2)}_{\ell s} 
        \left[
          D^{(j_1j_2)++}_{ss}
        + D^{(j_1j_2)--}_{ss}
        \right] 
        +
        \hatb^{(j_1)*}_{\ell s} \,\hatb^{(j_2)}_{\ell,s+2} 
        D^{(j_1j_2)+-}_{s,s+2}
        +
        \hatb^{(j_1)*}_{\ell,s+2} \, \hatb^{(j_2)}_{\ell s} 
        D^{(j_1j_2)-+}_{s+2,s}\right), 
\\
        & \simeq\frac{1}{4}
        \sum_{j_1 j_2}
        \hatb^{(j_1)*}_{\ell 0} \, \hatb^{(j_2)}_{\ell 0} 
        \left[
          D^{(j_1j_2)++}_{00}
        + D^{(j_1j_2)--}_{00}
        \right];
\\
{\cal W}^{EE}_{\ell} = {\cal W}^{BB}_{\ell} 
        &= \frac{1}{4}
        \sum_{j_1 j_2 s}\sum_{v=-2,2}\left(
        \hatb^{(j_1)*}_{\ell s} \, \hatb^{(j_2)}_{\ell s} 
        \left[
          D^{(j_1j_2)++}_{s+v,s+v}
        + D^{(j_1j_2)--}_{s+v,s+v}
        \right] 
        +
        \hatb^{(j_1)*}_{\ell s} \,\hatb^{(j_2)}_{\ell,s+2} 
        D^{(j_1j_2)+-}_{s+v,s+2+v}
        +
        \hatb^{(j_1)*}_{\ell,s+2} \, \hatb^{(j_2)}_{\ell s} 
        D^{(j_1j_2)-+}_{s+2+v,s+v}
        \right),
\\
        &\simeq \frac{1}{4}
        \sum_{j_1 j_2}
        \hatb^{(j_1)*}_{\ell 0} \, \hatb^{(j_2)}_{\ell 0} 
        \left[
          D^{(j_1j_2)++}_{22}
        + D^{(j_1j_2)--}_{22}
        + D^{(j_1j_2)++}_{-2-2}
        + D^{(j_1j_2)--}_{-2-2}
        \right];
\\
{\cal W}^{XY}_{\ell} &= 0 \quad \text{when $X \ne Y$}
%%%{\cal W}^{TB}_{\ell} = {\cal W}^{EB}_{\ell} = 0;
\end{align}
where the approximate results are obtained for circular beams. \\
Even for more realistic hypotheses on the hits locations,
the sub-pixel contributions to the respective power spectra follow the hierarchy
\begin{equation}
        \Delta \pC^{TT}_{\ell} \sim \Delta \pC^{EE}_{\ell} \sim \Delta \pC^{BB}_{\ell} 
                                \gg \Delta \pC^{TE}_{\ell} \sim \Delta \pC^{TB}_{\ell} \sim \Delta \pC^{EB}_{\ell}.
\end{equation}

Let us consider now the other extra contribution to the power spectrum, 
involving the Laplacian of the sky signal and the quadratic norm of $\dd\vecr$. 
Introducing
\begin{align}
        \cpem{j,v} \tomega_{v}^{(j)'}(p) &= \sum_{v'} (\matH^{-1}(p))_{vv'} \cpem{j,v'} \sum_{t \in p} 
        (\dd\theta_t^2 + \sin^2\theta_t \dd\varphi_t^2) f_{j,t} e^{i v' \alpha^{(j)}_t}\\
        &\simeq \sigma^2_p \cpem{j,v} \tomega_{v}^{(j)}(p)
\end{align}
where $\sigma^2_p$ is the second order moment of the hit location in pixel $p$. 
If we assume this and $\tomega_{v}^{(j)}(p)$
to be slowly varying functions of $p$, and consider $\ell \gg s$,
the power spectra become 
\begin{align}
C_\ell &\longrightarrow \left(1 - \frac{1}{2}\ell(\ell+1)\sigma^2\right) C_\ell
\end{align}
which describes to leading order, the smoothing effect of the integration of the signal on the pixel.
% \begin{align}
%       D^{(j_1j_2)'}_{v_1v_2, \ell'} &= 
%         \cpem{j_1,v_1}\cpem{j_2,v_2}\ 
%         \frac{1}{2\ell'+1}\sum_{m'}\left(
%         {}_{v_1}\tomega^{(j_1)}_{\ell'm'}
%         {}_{v_2}\tomega^{(j_2)'*}_{\ell'm'}+
%         {}_{v_1}\tomega^{(j_1)'}_{\ell'm'}
%         {}_{v_2}\tomega^{(j_2)*}_{\ell'm'}\right)
% \end{align}

\section{Co-polarized beam}
\label{sec:copolar_beams}
For an arbitrarily shaped beam having the intensity harmonics coefficients
\begin{align}
        b_{\ell m} &= \int \dd \vecr \bI(\vecr) Y_{\ell m}^*(\vecr),
\end{align}
and assumed to be perfectly co-polarized in direction $\gamma$, its polarized harmonics' content will be
\begin{align}
\ _{\pm 2}b_{\ell m} &= \int \dd \vecr \left(\bQ(\vecr)\pm i\bU(\vecr)\right)\ {}_{\pm2}Y^*_{\ell m}(\vecr)
                     = \int \dd \vecr\ \bI(\vecr)\ e^{\pm 2i (\gamma-\varphi_\vecr)}\  {}_{\pm 2}Y_{\ell m}^*(\vecr) \\
%
%       &= e^{\pm 2i \gamma} \sum_{\ell'm'}b_{\ell'm'} \int \dd \vecr\ e^{\mp 2i \varphi_\vecr}\  Y_{\ell'm'}(\vecr)\ {}_{\pm 2}Y_{\ell m}^*(\vecr) \\
        &= e^{\pm 2i \gamma} \sum_{\ell'm'}b_{\ell'm'} 
        \int_0^{2\pi} \dd\varphi 
        \int_0^{\pi}\dd\theta \sin\theta 
        \ e^{\mp 2i \varphi_\vecr}\  Y_{\ell'm'}(\theta,\varphi)\ {}_{\pm 2}Y_{\ell m}^*(\theta,\varphi) \\
        &= e^{\pm 2i \gamma} 2\pi \sum_{\ell'}b_{\ell',m\pm 2}  (-1)^m 
        \sum_{\ell'' \ge 2} \left(\frac{(2\ell+1)(2\ell'+1)(2\ell''+1)}{4\pi} \right)^{1/2}
        \wjjj{\ell}{\ell'}{\ell''}{-m}{m \pm 2}{\mp 2} 
        \wjjj{\ell}{\ell'}{\ell''}{\pm 2}{0}{\mp 2}\  I_{\ell''}
        \label{eq:blm_polar_3J}
\end{align}
where we used Eq.~(\ref{eq:Ylm_product}) and introduced, for $\ell'' \ge 2$
\begin{align} %Mathematica
I_{\ell''} &\equiv \int_0^{\pi} \dd\theta \sin\theta \  _{\pm 2}Y_{\ell'',\mp 2}^*(\theta),\\
        &= \sqrt{\frac{2\ell''+1}{4\pi}} \int_0^{\pi} \dd\theta \sin\theta\  d^{\ell''}_{\mp2,\mp2}(\theta),\\
        &= \sqrt{\frac{2\ell''+1}{4\pi}} \frac{4}{\ell''(\ell''+1)} (-1)^{\ell''}.
\end{align}
Since $I_{\ell''}$ peaks at $\ell''=2$, the 3J symbols will enforce $\ell'' \ll \ell \simeq \ell'$ in Eq.~(\ref{eq:blm_polar_3J}).
If the beam is narrow enough in real space, $b_{\ell',m}$ will be almost constant over the allowed range 
$\ell-\ell'' \le \ell' \le \ell+\ell''$, and we use
 the approximate orthogonality relation of Eq.~(\ref{eq:w3j_ortho3}) to write
\begin{align}
        \sum_{\ell'}b_{\ell',m\pm 2} (-1)^m \sqrt{(2\ell+1)(2\ell'+1)}
        \wjjj{\ell}{\ell'}{\ell''}{-m}{m \pm 2}{\mp 2} 
        \wjjj{\ell}{\ell'}{\ell''}{\pm 2}{0}{\mp 2} 
        &\simeq b_{\ell,m\pm 2} \sum_{\ell'} (-1)^m  (2\ell'+1)
%       \wjjj{\ell}{\ell'}{\ell''}{-m}{m \pm 2}{\mp 2} 
        (-1)^{m \pm 2}
        \wjjj{\ell}{\ell'}{\ell''}{\pm 2}{0}{\mp 2} ^2, \nonumber \\
        &=b_{\ell,m\pm 2}.
\end{align}
Finally, we note that in Eq.~(\ref{eq:blm_polar_3J}), the sum
\begin{align}
        2\pi \sum_{\ell''=2}^{2n+1}  \sqrt{\frac{2\ell''+1}{4\pi}} I_{\ell''} 
&=
        \sum_{\ell''=2}^{2n+1} (-1)^{\ell''} \frac{2(2\ell''+1)}{\ell''(\ell''+1)},  \\
&= \sum_{p=1}^{n}\frac{1}{p(p+1)}
= \sum_{p=1}^{n}\frac{1}{p}-\frac{1}{p+1}, \\
&= 1 - \frac{1}{n+1},
\end{align}
to obtain
\begin{align}
\ _{\pm 2}b_{\ell m}&= e^{\pm 2i \gamma}b_{\ell,m\pm 2},
\end{align}
which is valid for any (narrow) co-polarized beam.

%\end{widetext}

\end{document}